\documentclass[pdftex,useAMS,usenatbib]{mn2e}
\usepackage{graphicx}
\usepackage{multirow}
\usepackage{ifthen}
\usepackage{hyperref}
\usepackage{stfloats}
\usepackage{fixltx2e}
\usepackage{color}

\newcommand{\msun}{${\rm M_{\sun}}$}

\def\ltsima{$\; \buildrel < \over \sim \;$}
\def\simlt{\lower.5ex\hbox{\ltsima}}
\def\gtsima{$\; \buildrel > \over \sim \;$}
\def\simgt{\lower.5ex\hbox{\gtsima}}
\def\kms{{\rm\,km\,s^{-1}}}

\def\kpc{{\rm\,kpc}}

\def\msun{{\rm\,M_\odot}}

\makeatletter
\newcommand\ion[2]{#1$\;${\small\rmfamily\@Roman{#2}}\relax}% 
\makeatother
\def\s{\ifmmode \widetilde \else \~\fi}
\def\={\overline}

\def\spose#1{\hbox to 0pt{#1\hss}}

\def\lta{\mathrel{\spose{\lower 3pt\hbox{$\mathchar"218$}}
     \raise 2.0pt\hbox{$\mathchar"13C$}}}
\def\gta{\mathrel{\spose{\lower 3pt\hbox{$\mathchar"218$}}
     \raise 2.0pt\hbox{$\mathchar"13E$}}}
\def\Dt{\spose{\raise 1.5ex\hbox{\hskip3pt$\mathchar"201$}}}    % upper case
\def\dt{\spose{\raise 1.0ex\hbox{\hskip2pt$\mathchar"201$}}}    % lower case

\def\dotsfill{\leaders\hbox to 1em{\hss.\hss}\hfill}

\loadboldmathitalic 
\title[Stellar Streams as Probes of Dark Halo Mass and Morphology]
  {Stellar Streams as Probes of Dark Halo Mass and Morphology: A Bayesian Reconstruction}
\author[A. Varghese, R. Ibata, G. F. Lewis]
 {A.~Varghese,$^{1}$\thanks{E-mail: anjali.varghese@astro.unistra.fr}
  R.~Ibata,$^{1}$ G. F.~Lewis,$^{2}$  \\
  $^{1}$Observatoire Astronomique de Strasbourg (UMR 7550),
      11, rue de l'Universit\'e, 67000 Strasbourg, France\\
$^{2}$Sydney Institute for Astronomy, School of Physics, A28, University of Sydney, NSW, 2006 Australia}

\date{\today}
\begin{document}
\maketitle 
\begin{abstract}
Tidal streams provide a powerful tool by means of which the matter distribution of the dark matter halos of their host galaxies can be studied. However, the analysis is not straightforward because streams do not delineate orbits, and for most streams, especially those in external galaxies, kinematic information is absent. We present a method wherein streams are fit with simple corrections made to possible orbits of the progenitor, using a Bayesian technique known as Parallel Tempering to efficiently explore the parameter space. We show that it is possible to constrain the shape of the host halo potential or its density distribution using only the projection of tidal streams on the sky, if the host halo is considered to be axisymmetric. By adding kinematic data or the circular velocity curve of the host to the fitting data, we are able to recover other parameters of the matter distribution such as its mass and profile. We test our method on several simulated low mass stellar streams and also explore the cases for which additional data are required.
\end{abstract}


\section{Introduction}
\label{Introduction}
The global features of the dark matter halos of spiral galaxies such as their mass, shape and extent hold key clues to understanding the nature of dark matter particles as well as galactic evolution and morphology. Simulations of cold dark matter tend to produce halos that are oblate \citep{Dubinski:1994p1963, Debattista:2008p3979} or triaxial (\citealt{Frenk:1988p3563, Dubinski:1991p3573, Bett:2007p3576}) whereas hot dark matter models yield more spherical halos \citep{Mayer:2002p2795, Bullock:2002p4128} and baryonic dark matter such as cold molecular gas form disk-like halos \citep{Pfenniger:1994p1978}. It is crucial to test these theoretical predictions with observations and to this end, several techniques have been developed to probe the vertical distribution of dark matter around galaxies. These range from methods that are very local, such as measuring the density of dark matter in the Solar neighbourhood which is sensitive to the halo flattening \citep{Kuijken:1989p1968}, to methods based on gravitational lensing \citep{Hoekstra:2004p4990, Mandelbaum:2006p5128}. Yet, a conclusive result remains elusive. This is in part due to the small sample of galaxies for which these measurements have been made. The other cause for the disagreement in results may be the systematic differences in the methods themselves. For instance, \citet{Olling:2000p5209} measured the flattening of halos based on its effect on the thickness of the \ion{H}{1} gas layer in disk galaxies and find that their method tends to yield flatter halos as opposed to other techniques such as those based on warped gas disks. A recent study of halo shapes based on \ion{H}{1} flaring can be found in \cite{OBrien:2010p4026, OBrien:2010p4024, OBrien:2010p4015, OBrien:2010p4032}. See \cite{Sackett:1999p5285} for a review of earlier work on the subject, and \cite{Merrifield:2004p5320} for measurements of the shape of the Milky Way halo. \cite{Allgood:2006p4765} contains a more recent comparative study of different methods to determine halo shapes. 

Orbits in a potential are simple, yet powerful tracers of the matter distribution. Polar rings in early-type galaxies are a good example of these \citep{Sackett:1994p2525, Iodice:2003p4379, Combes:2006p4362}. In this paper, we use a similar tracer, namely the stellar streams that are formed by the tidal disruption of globular clusters or dwarf spheroidal galaxies as they fall into the halos of larger spiral galaxies. The stars that make up the stream are ejected from the in-falling satellite due to tidal disruption and shocks during pericenter passages. The stream stars occupy (approximately) two different orbits that are slightly offset from that of the progenitor, one that is ahead of it and one lagging behind, making up the leading and trailing arms of the stream respectively. The positions of the stream stars can be calculated fairly easily from the orbit of the progenitor (see \S\ref{streamfitting}). As the geometry of the stream depends on the orbit of the progenitor, which in turn depends on the potential of the host galaxy, it can be used to constrain
the parameters of the halo potential or its density distribution. Tidal streams in the halo have the added advantage of being far enough from the bright galactic components that the fitting is unaffected by small errors in the disk and bulge models. Tidal tails formed by interacting disk galaxies in major mergers can also be used to probe the parent halos \citep{Dubinski:1999p4497, Springel:1999p4494}, but in this study, we focus solely on stellar streams formed in minor mergers. 

Several stellar streams have been observed in the Milky Way; these include the Sagittarius \citep{Ibata:2001p2812, Majewski:2003p5080}, the Orphan \citep{Belokurov:2007p4658, Grillmair:2006p3088} and the Monoceros tidal streams \citep{Newberg:2002p3260, Conn:2005p3233, Conn:2007p3172, Conn:2008p3209}.  Stream-like structures have also been observed around many nearby galaxies, for example M31 \citep{Ibata:2001p4924, Ibata:2004p4927}, NGC 5907 \citep{MartinezDelgado:2008p387}, NGC 891 \citep{Mouhcine:2010p3269} and others; refer \citet{MartinezDelgado:2010p2136} for a recent systematic survey of streams in nearby spiral galaxies. Many more remain to be uncovered by future surveys that are increasingly sensitive to low surface brightness objects, enabling an extensive application of techniques based on them. These could provide a large sample of measurements of halo shapes and profiles, which is required if we are going to be able to uncover the generic properties of dark halos and study the possible correlations with their formation histories. 

There has been an ongoing effort in recent years to glean information from these streams about their host halo distribution. For example, the fact
that the stars of the Sagittarius tidal stream lie on a narrow great circle on the sky (implying little precession) and that the stream has not been
dispersed, initially led to the conclusion that the Milky Way has a spherical halo \citep{Ibata:2001p2812, Majewski:2003p5080}.  \citet{Mayer:2002p2795}, \cite{Helmi:2004p4963} and \citet{MartinezDelgado:2002p2825} suggested that this need not necessarily be the case as the debris is dynamically young and may not be sensitive to the shape of the halo. \cite{Law:2009p2119} have shown that the Sagittarius tidal stream is best reproduced in a triaxial halo and more recently, a more mildly triaxial halo \citep{Law:2010p5482}. Attempts have also been made to constrain the Galactic potential by fitting thin, cold streams found in the Milky Way halo, GD1 \citep{Grillmair:2006p5594} and Palomar 5 \citep{Odenkirchen:2003p5606}, with orbits integrated in a triaxial potential, but the data were found to be insufficient to discriminate between possible solutions, although triaxial models seem to be favored over spherical ones \citep{Lux:2010p5614}. However, by fitting the GD-1 stream with orbits in a logarithmic, axisymmetric potential, \citet{Koposov:2010p6012} have estimated the flattening of the Milky Way potential at galactocentric radii near $R \sim15 \kpc$ at $0.9$ and a lower limit for the flattening of its halo potential at $0.89$.

In general, there are two main challenges that have so far hindered the successful application of tidal streams in constraining halo shapes. One is that tidal streams do not exactly delineate individual orbits in the galactic potential (\citealt{Eyre:2009p5036, Eyre:2009p5033}) and treating them as such may result in incorrect estimates. We overcome this hurdle by fitting a given stream with corrected sets of points computed from the progenitor orbit. The other limitation is that in most systems, only projected coordinates of the stream on the sky are available. In closer systems such as the Andromeda galaxy (M31), line of sight velocities are also available and distances to globular clusters or dwarf galaxies that may be the stream progenitors are measurable using the tip of the Red Giant Branch (TRGB method - see \citealt{McConnachie:2004p5175}. In previous studies using tidal streams, the approach has been to try to reproduce the observed streams using a few N-body simulations (typically less than 10). It has been found that the phase space information is insufficient in reasonably constraining the halo parameters using this approach. The N-body technique is severely handicapped by the fact that it is only possible to explore a tiny fraction of the full parameter space with such simulations. We redress this shortcoming by adopting a statistical approach sampling the parameter space of possible progenitor orbits and halo parameters using a Markov chain Monte Carlo routine. The sampling yields distributions of the halo structural parameters that peak at their most likely values. We find in our tests on simulated streams that it is possible to estimate halo shapes using this method even with limited phase space information. We also find that it is possible to recover line of sight distances along streams in cases where they are not available. In this paper, we explore how much and what information is required to uniquely estimate the orbital and potential parameters using this statistical approach and how the estimates improve with additional data. Later contributions in this series will use the technique developed here to actual observed systems.

\section{Methodology}
Consider a galaxy whose disk lies in the $XY$ plane. We assume that the mid-plane of its bulge and halo coincide with this plane. We also assume that the halos we study are static and have an axisymmetric density profile, even though the density profile of the dark halo is generally considered to be a triaxial ellipsoid (\citealt{Dubinski:1991p3573, Jing:2002p4163, Bailin:2005p4487}). The flatness of the distribution is given by the ratio of the polar to the equatorial axis $(c/a)$ and the ovalness by the equatorial axis ratio $(b/a)$. In the most general models, which involve a superposition of ellipsoids, $c/a$ and $ b/a$ vary with radius. 

However, for the present contribution, we make the simplifying assumption that the equatorial axes are equal $(b/a = 1)$,  and that the halo has a flattening of $q = c/a$ (we discuss both flattening in the potential and density distribution). Studies have shown that this is a reasonable assumption for the halos of spiral galaxies, especially far from the galactic disk where it becomes spheroidal (\citealt{Debattista:2008p3979, Kazantzidis:2010p3520}). 

Throughout the paper, we analyse the specific case of a host galaxy that is viewed edge-on, although the method can be very simply adapted to systems that do not have that geometry. What we have access to are the projected coordinates ($x$,$z$) of a tidal stream in its potential, and the width of the stream providing an uncertainty of $\sigma$ on these coordinate values. The tidal stream has two tails: the leading and the trailing tails. In reality, each tail consists of several different yet very similar orbits, the stars along these having similar values of energy and angular momentum. The slight difference in these orbits show up in places in the form of bifurcations and small protrusions from the main stellar feature. In this paper, we neglect these tiny features and only consider the longest, contiguous structures that we observe. The method is only applicable in its current simple form to streams of low mass satellites ($\simlt 10^8 \msun$), for which the self-gravity of the stream is negligible. This allows us to ignore dynamical friction which would have to be taken into consideration for more massive and heavily disrupted systems.

The method for fitting the streams is briefly as follows. Different values of the halo potential parameters and the orbital parameters of the progenitor are tested, each set of values corresponding to a point in the multidimensional parameter space. For each set of parameters considered, an orbit is integrated and a set of points is calculated that represents the stream that would be formed by a satellite on the orbit (see \S\ref{streamfitting}); we refer to this as the trial stream corresponding to the point in parameter space under consideration. Comparing the trial stream to the observed stream (or the N-body generated test stream in the present work), we measure how likely the parameters are to be the ones that generated the observed stream. We sample different values of the parameters using a Markov chain Monte Carlo (MCMC) algorithm, in order to deduce the distribution of parameter values that are consistent with the data. For the galactic halo potential, we use two models: a purely logarithmic halo for its simplicity and a more realistic multiple-slope power-law model.

In \S\ref{Testing}, we explore how the quality of the estimate depends on the quantity and type of information available. We discuss cases for which the streams are shorter or have fewer turning points. We also consider the different kinds of data that may be available and their effect on the accuracy of the estimate. For instance, line of sight velocities and/or distances at some points along the stream or the rotational velocity curve are available for nearby systems. The method is tested on these various cases with pseudo-data generated by N-body simulations.

\section{The Fitting Routine}
\label{OrbitalFitting}

In this section, we first briefly discuss the general principles of maximum likelihood estimation and parallel tempering, and then describe how we apply it to fitting tidal streams. 

Maximum likelihood parameter estimation is a robust method of determining the parameters of a model that maximize the likelihood of a data set. The likelihood of a set of parameters is the probability of obtaining the given data for those parameter values. Suppose the data set consists of $N$ independent observations of a random variable $x$ which has a probability distribution function $f(x;\Theta)$ that depends on a set of $k$ parameters $\Theta$ which are to be estimated, then the likelihood is calculated by the product of the values of the distribution function for each data point \citep{Gregory:2005p5433}:
\begin{equation} \label{eq:likelihood}
\mathcal{L}(x_{1},x_{2},...,x_{N}|\Theta_{1},\Theta_{2},...,\Theta_{k}) = \prod_{i=1}^{N} f(x_{i};\Theta_{1},\Theta_{2},...,\Theta_{k}).
\end{equation}
The estimated values of the parameters $\Theta$ are the ones for which this likelihood is maximum. A basic and simple algorithm to find this set of parameters is known as the Metropolis Algorithm, which probes the parameter space in an efficient way to locate the parameters for which the likelihoods are high. The sampling is done by a chain that walks through the parameter space.

For a single chain, the walk through the parameter space basically involves choosing a random point that we consider to move to, and then moving to it or staying put based on a likelihood criterion. Such a chain is known as a Markov chain Monte Carlo (MCMC). A chain is set off at an initial random point $\Theta_{t = 0}$ in the parameter space (t being the time step of the MCMC), and we calculate the likelihood at this point, say $\mathcal{L}(\Theta_{t})$. Then we consider another point $\Theta'$ (the trial point) in the parameter space, from a proposal distribution and calculate its likelihood $\mathcal{L}(\Theta')$. In our algorithm, the proposal distribution is a normal distribution centered on the current point. If the trial point falls beyond the plausible range of the parameters, then we set its likelihood to be arbitrarily low. We calculate the Metropolis ratio, $r = \mathcal{L}(\Theta_{t})/\mathcal{L}(\Theta')$. If this is greater than 1, then we take the trial point as the next point in the chain. If $r <1$, then we pick a uniform random number $U$ between 0 and 1. If $U \leq r$, then we take the trial point as the next point ($\Theta_{t+1} = \Theta'$), else we stay at the same point ($\Theta_{t+1} = \Theta_{t}$). Proceeding in a similar fashion, we explore the full parameter space. 

\subsubsection{Parallel Tempering}
In many cases, especially in multidimensional parameter space, there are usually many local likelihood maxima, whereas we seek the global maximum. Using only a single chain, the algorithm often becomes stuck in a local maximum. To avoid this, multiple chains of different ``temperatures'' are used to step through the parameter space. Effectively, this means that for a chain of temperature $T$, its likelihood is raised to the power of the inverse of its temperature, i.e.:
\begin{equation} \label{eq:lbeta}
\mathcal{L}_{\beta_{i}}= (\mathcal{L}_{1})^{\beta_{i}} \, ,
\end{equation}
\noindent where $\mathcal{L}_{1}$ corresponds to the likelihood at a temperature of 1 (given by equation \ref{eq:likelihood} in general, which reduces to equation \ref{eq:like} for the present study) and $\beta_{i} = 1/T_{i}$.  $T_{i}$ is the temperature of the  $i^{th}$ chain. As $T$ ranges from 1 to infinity, $\beta$ takes values from 1 to 0. The higher temperature chains are less sensitive to the likelihood, and take larger strides through the parameter space. Suppose we use $n$ parallel chains and swap their states after every $n_{s}$ steps on average. The higher temperature chains pull out the colder chains from local maxima they may be stuck in.  The algorithm for the swapping of chains is as follows: 
\begin{enumerate}
\item At each step, choose a uniform random number $U_{1}$ between $0$ and $1$. A swap is proposed if $U_{1}  \leq 1/n_{s}$.
\item If a swap is proposed, we consider swapping the states of chain $i$ and $i+1$, where $i$ is a random integer between $1$ and $n-1$.
\item Calculate the swapping probability,
\begin{equation}
r = \frac{ \mathcal{L}_{\beta_{i}}(\Theta_{t,i+1} ) \mathcal{L}_{\beta_{i+1}}(\Theta_{t,i})} {\mathcal{L}_{\beta_{i}}(\Theta_{t,i} ) \mathcal{L}_{\beta_{i+1}}(\Theta_{t,i+1})}
\end{equation}
\hspace*{4 mm} where $\Theta_{t,i}$ and $\Theta_{t,i+1}$ are the current points on the $i^{th}$ and $(i+1)^{th}$ chains respectively, and $\mathcal{L}_{\beta_{i}}(\Theta_{t,i+1})$ is the likelihood of $\Theta_{t,i+1}$ if it were on the $i^{th}$ chain.
\item Choose another uniform random number $U_{2}$ between 0 and 1. Accept the swap if $U_{2} \leq r$.
\item The maximum corresponding to the coldest chain $T = \beta = 1$ is the global maximum.
\end{enumerate}
The efficiency of the algorithm is sensitive to the size of the proposal distribution. If it is too small, then most of the trial points are accepted and the MCMC is slow to sample the full parameter space. If it is too large, most of the trial points are rejected and though the MCMC may make large jumps in parameter space, it could become stuck at a certain point for long. For the present work, optimal sizes of the proposal distributions for the parameters were found experimentally, by requiring that the acceptance rate lie between $15\%$ and $40\%$. Note that it is possible, however, to automate the process such that the routine finds an optimal size for the proposal distribution \citep{Gregory:2005p5433}. It is important to check that the MCMC converges on a solution. One way of checking for convergence is to sample the parameter space with chains that start at different initial points and to check if they yield similar solutions. Once convergence has been achieved (i.e. a chain is well-mixed), the distribution is independent of the number of MCMC steps and is said to be stationary. The initial steps of the chain are discarded to forget the starting point. This is known as burn-in and is typically around $10,000$ steps in our test cases. The remaining steps of the chain form a sample distribution of the parameters. The estimated value of a parameter corresponds to the peak of its marginalized distribution. The accuracy of the estimate is equal to the standard deviation of the distribution, if it turns out to approximate a Gaussian distribution. However, the parameter distributions may turn out to be multimodal or otherwise complex. As the specific details of the parameter distributions vary from stream to stream, we defer the discussion of accuracies and convergence tests to when we apply the method to real, observed streams. For a detailed discussion of Parallel Tempering and maximum likelihood estimation, see \cite{Gregory:2005p5433}.

\subsection{Estimation of Halo Parameters}
By defining the parameter space and the likelihood calculation for the parameters, we can use the above method to find an orbit that best reproduces a given stream (after the required corrections explained in \S\ref{streamfitting}), thereby also estimating the model parameters that we wish to determine. A point in the parameter space corresponds to an orbit, which is parametrized by its initial position and velocity components ($x_{0}$, $y_{0}$, $z_{0}$, $v_{x_{0}}$, $v_{y_{0}}$, $v_{z_{0}}$ at $t = 0$), in addition to the potential or density distribution parameters. Depending on the information available, few of these initial phase space coordinates are known. The remaining orbital parameters are set to be free parameters with physically reasonable ranges. The best fit orbit corresponds to the one with the highest likelihood. The main processes involved in the algorithm are: (1) Stepping through the parameter space, (2) Integrating orbits corresponding to the points in the parameter space, (3) Generating a corrected set of points (trial stream) corresponding to an orbit and (4) Calculating the likelihood of these trial streams. 

As the several MCMC chains step through the parameter space, we integrate orbits at each step with a fourth-order Runge-Kutta  integrator starting from an initial point defined by the parameters. For fitting the streams, we calculate a set of $n$ points $\{S(n)\}$ via a correction mechanism (detailed in \S\ref{streamfitting}) from the integrated orbit. Assuming a Gaussian distribution for the parameters, the likelihood of a trial orbit is calculated as:
\begin{equation} \label{eq:like}
\mathcal{L} = \prod_{i=1}^{N} \frac{1}{\sigma_{i} \sqrt{2\pi}} e^{-\frac{D_{i}^{2}}{2\sigma_{i}^{2}} } \, ,
\end{equation}
where $N$ is the number of data points and $\sigma_{i}$ is the uncertainty in each data point, which includes both the width of the stream as well as measurement errors. $D_{i}$ is the distance of each data point on the observed stream from the corrected set of points, i.e. the distance of a data point to its closest point on the trial stream. As not all of the phase space information is available, the distance $D_{i}$ only includes the distances along the coordinates for which data are available. For instance, if the projected positions $x_{i}, z_{i}$ and line of sight (l.o.s) velocities $v_{y_{i}}$ are observed, then the exponent in equation \ref{eq:like} reduces to:
\begin{equation} \label{eq:dist}
\frac{D_{i}^{2}}{2\sigma_{i}^{2}} = \sum_{i} \frac{(x_{i} - x_{c})^{2} + (z_{i} - z_{c})^{2}}{2\sigma_{xz_{i}}^{2}} + \frac{(v_{y_{i}} - v_{y_{c}})^{2}}{2\sigma_{v_{i}}^{2}}\, ,
\end{equation}
where $x_{c}, z_{c}, v_{y_{c}}$ are the closest points on the trial stream to the data point $x_{i}, z_{i}, v_{y_{i}}$ and $\sigma_{xz_{i}}, \sigma_{v_{i}}$ are the uncertainties in position and velocity. The $\sigma_{i}$ in the normalizing coefficient of the exponent in equation \ref{eq:like} is the product of the uncertainties in the different phase space coordinates, which in this example reduces to $\sigma_{xz_{i}} \sigma_{v_{i}}$.

In our tests, we used four parallel chains to search the parameter space, the inverse of the temperature of the $i^{th}$ chain being $\beta_{i} = 1/i^{2}$. The states of the chains are considered for swapping after every thirty steps on average. At the beginning of the parallel tempering, all the chains are initialized by setting the free parameters to physically plausible random values. We checked that the results were insensitive to these starting values. Given the stream data, the likelihood of the orbit corresponding to the initial point is calculated with equation \ref{eq:like}. This is the likelihood of the starting point for the coldest chain, $i=1$. The likelihoods for the same point on the hotter chains are calculated using equation \ref{eq:lbeta}.  

Having calculated the likelihoods for the initial point, the next point on each MCMC has to be chosen. For this, a trial point is chosen for each chain. This can be done in many ways. In our case, if $\Theta_{t}$ is the current value of the parameter $\Theta$, then the next point for consideration $\Theta'$ is drawn randomly from a normal distribution centered on $\Theta_{t}$. The width of the proposal distribution, \emph{i.e.} the size of the step, depends on the expected range of the parameters. It also helps to use different step sizes for the different chains to enable sampling the parameter space at various scales. The algorithm is run until the chains are well mixed, which could take from approximately one hundred thousand to a million steps (depending on the stream and halo model), and the marginalized distributions of the parameters are drawn from the coldest chain.

\section{Testing with Orbits (not streams)}
\label{Testing}

To get a heuristic idea of the effectiveness of the method, we first investigated how well it constrains potential parameters using ideal orbits in both logarithmic halos and double power law density halos. The primary aim was to explore if and when the projected positions of an orbit are sufficient to estimate the density or potential flattening and how additional information would affect the accuracy of the estimation.

\subsection{Orbits in a Logarithmic Potential}
\label{logorbits}

\begin{figure}
\begin{center}
 \includegraphics[width = 90mm, height = 80mm]{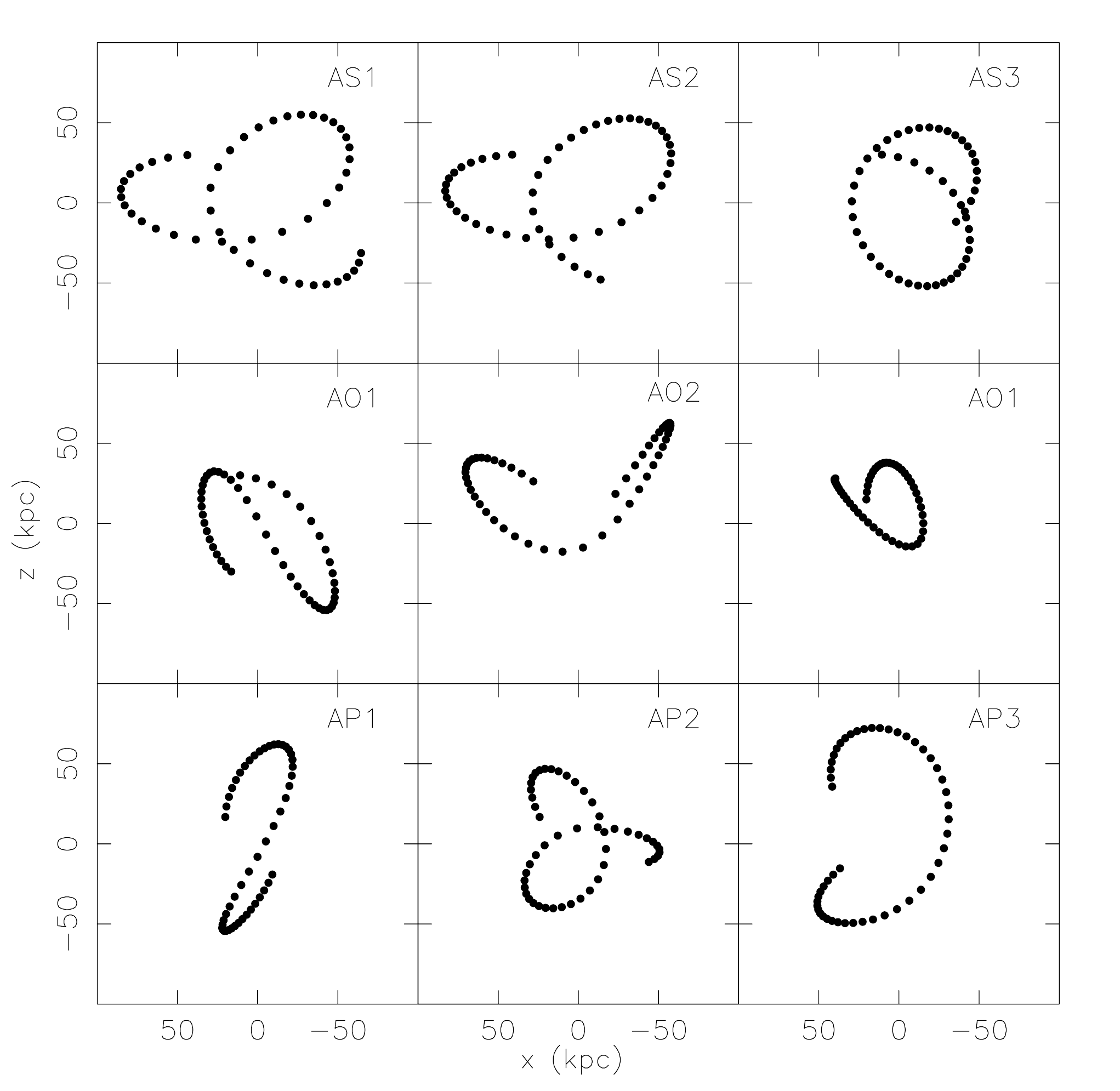}
\caption{Projection in the $XZ$ plane of orbits integrated in a logarithmic potential using a Runge-Kutta scheme. The dots indicate the positions on the orbits that are used as data points in the fitting. The top, middle and bottom panels show orbits in spherical, oblate and prolate potentials respectively. The parameters of each orbit are listed in Table \ref{tablelogseries}.}
\label{figlogseries}
\end{center}
\end{figure}

\begin{figure}
\begin{center}
 \includegraphics[width = 80mm, height = 80mm]{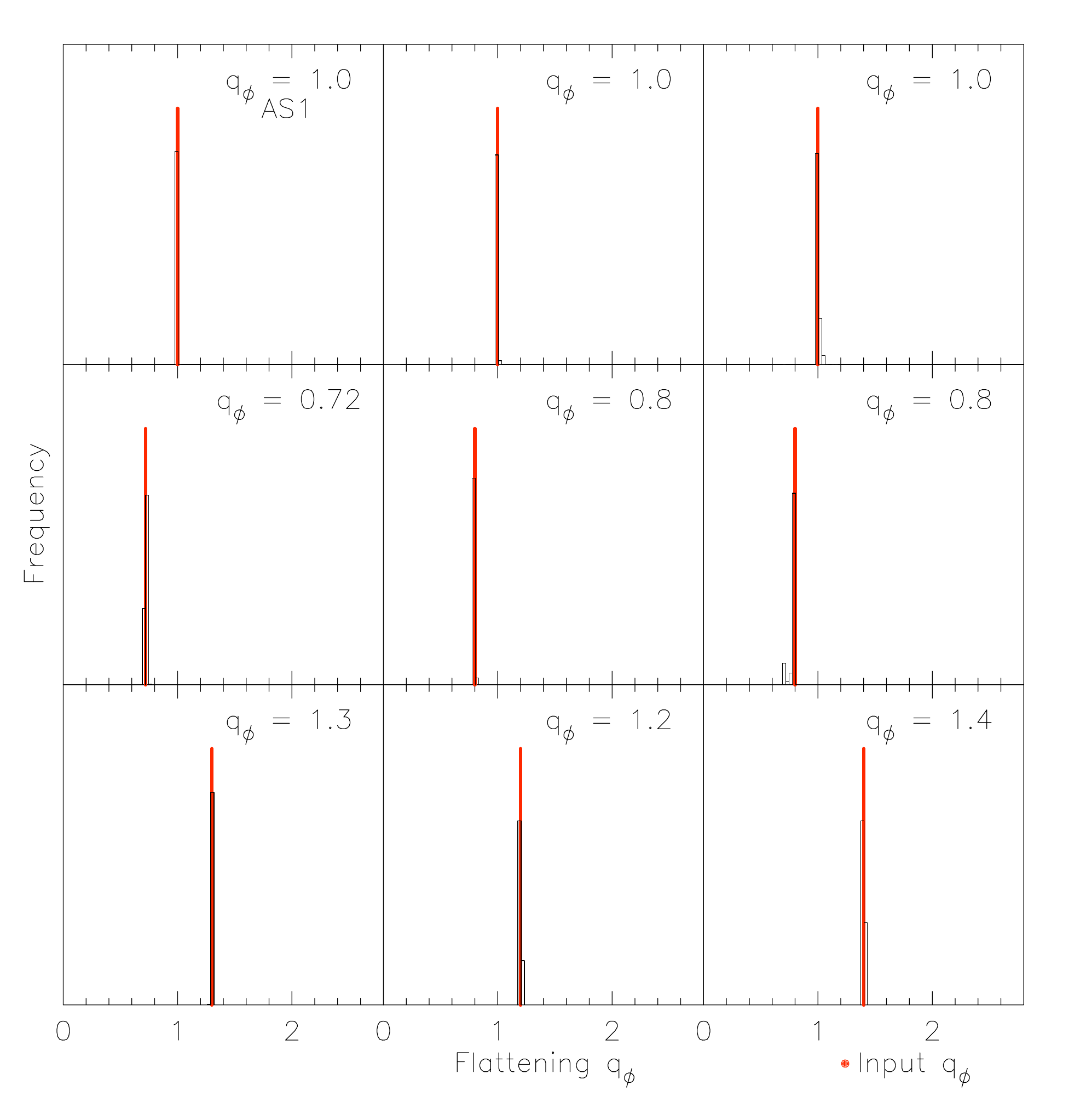}
\caption{Estimation of $q_{\phi}$ for the projected orbits shown in Figure \ref{figlogseries}. The input value of $q_{\phi}$ in each case is shown. This distribution is drawn from 100,000 steps of the coldest MCMC chain. The excellent correspondence with the input values shows that the shape of a logarithmic potential can be accurately recovered from the spatial projection of some orbits within it.}
\label{figqplots}
\end{center}
\end{figure}

\begin{figure}
\begin{center}
 \includegraphics[width = 80mm, height = 80mm]{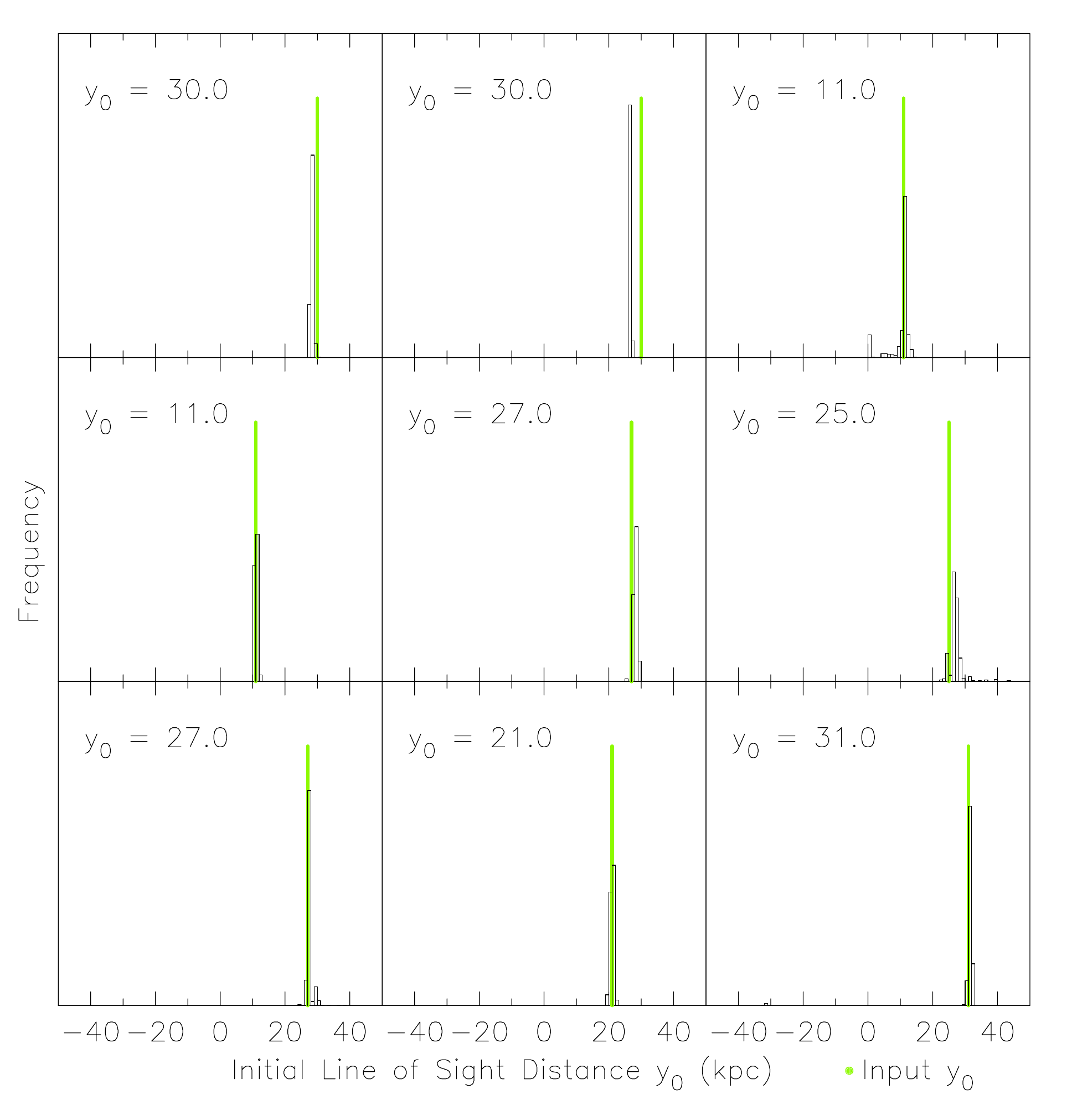}
\caption{Estimation of $y_{0}$ for the projected orbits shown in Figure \ref{figlogseries}. The input value of $y_{0}$ in each case is shown. This distribution is drawn from 100,000 steps of the coldest MCMC chain, and clearly peaks near the input value. The fitting routine only considers positive values for $y_{0}$ as orbits that are symmetrical about the $XZ$ plane will have identical positions (so there may be a sign discrepancy in the estimated value of $y_{0}$).}
\label{figyplots}
\end{center}
\end{figure}

\begin{figure}
\begin{center}
 \includegraphics[width = 80mm, height = 80mm]{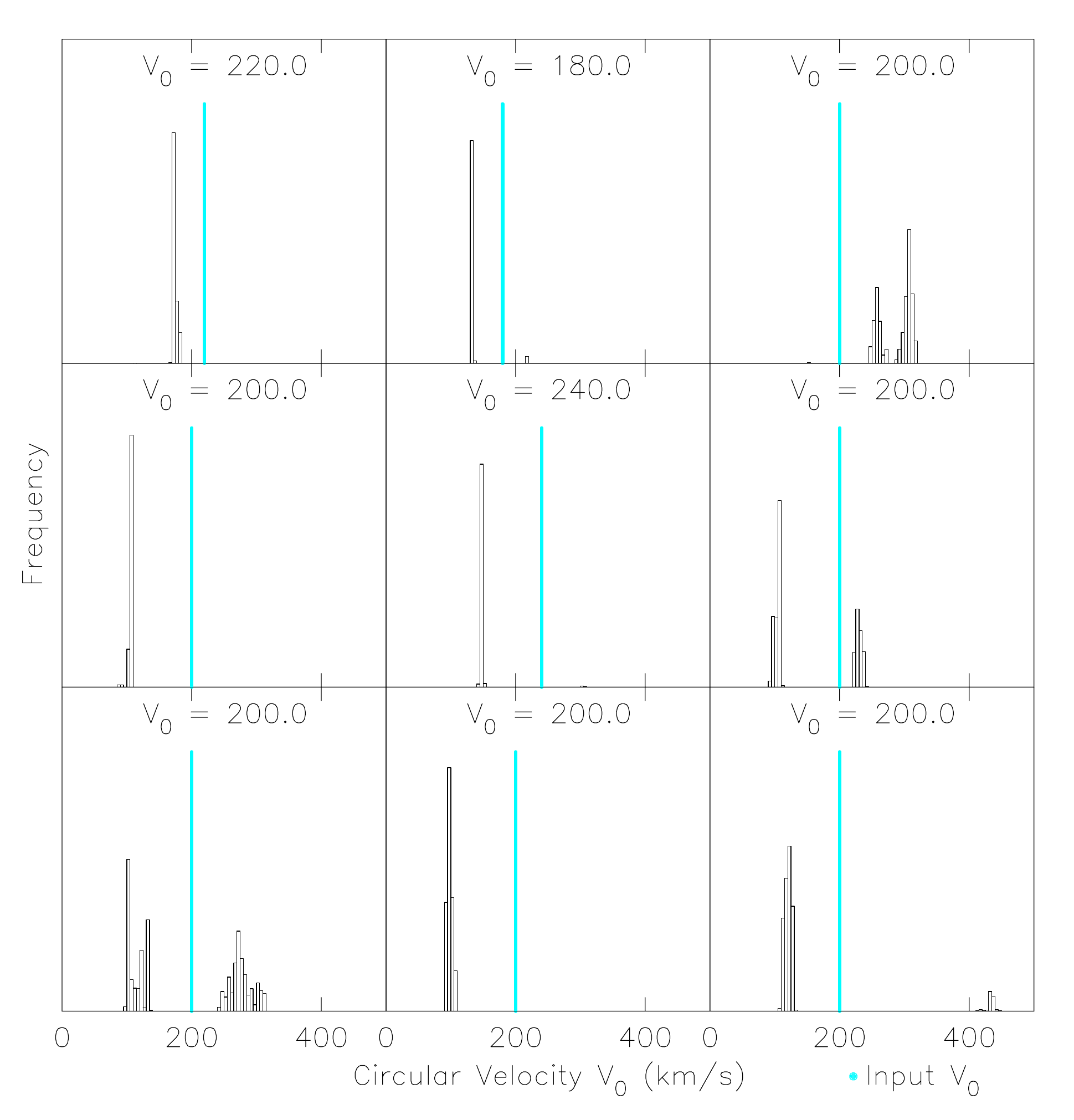}
\caption{Estimation of $V_{0}$ for the orbits shown in Figure \ref{figlogseries}. The input value of $V_{0}$ in each case is shown. This distribution is drawn from 100,000 steps of the coldest MCMC chain. We see that the peaks of the distribution vary greatly from the input value, which implies that the orbits are not uniquely dependent on the circular velocity.}
\label{figvcplots}
\end{center}
\end{figure}

\begin{figure}
\begin{center}
 \includegraphics[width = 80mm, height = 80mm]{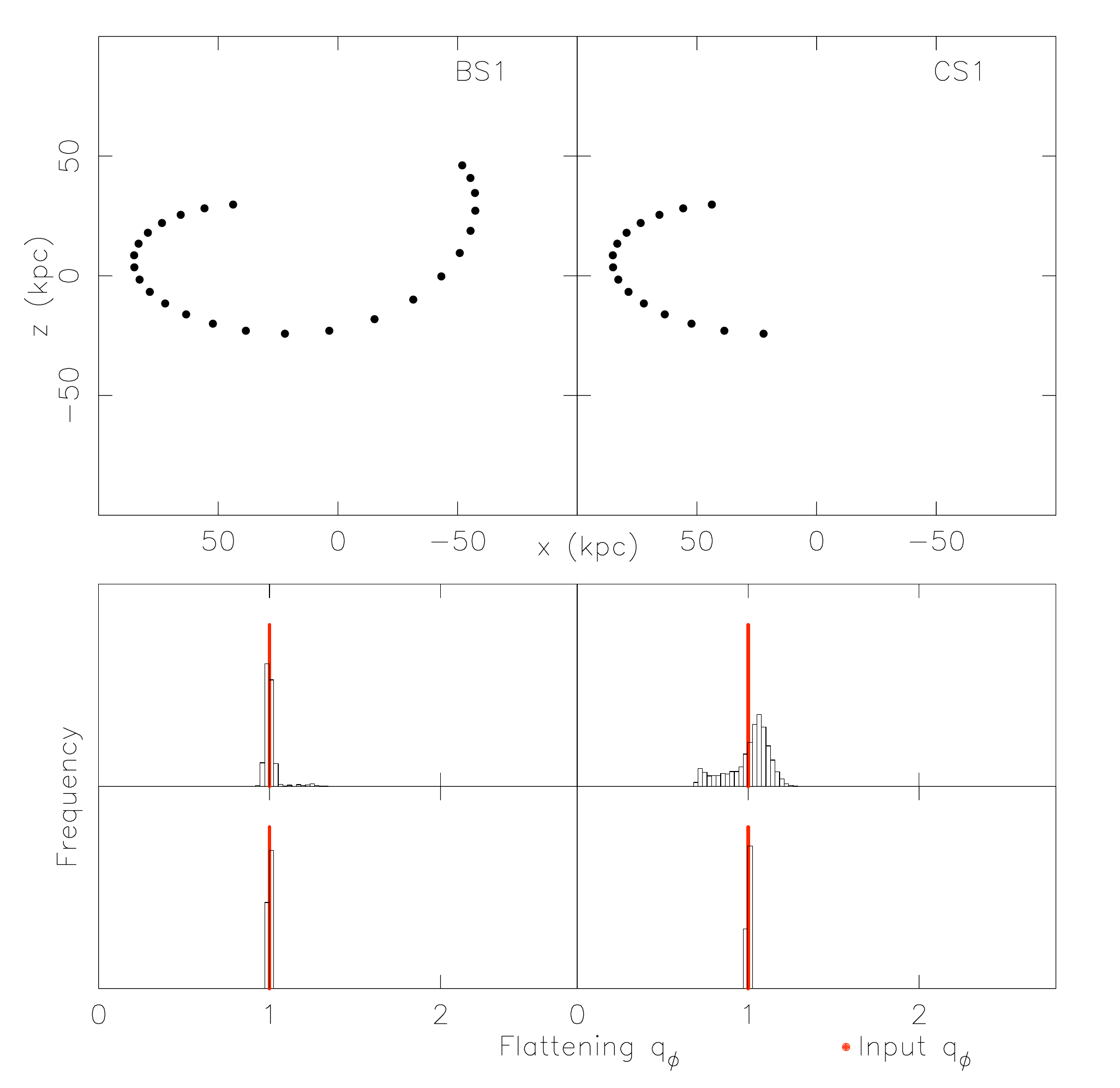}
\caption{Fitting shorter versions of the orbit AS1 ($q_{\phi}=1$). BS1 has two turning points and CS1 only one. The middle panels show the $q_{\phi}$ distributions obtained by fitting only the projected positions of these orbits. The spread in the distribution increases with decreasing turning points. The bottom panels show the $q_{\phi}$ distribution obtained by adding more information to the fitting: the line of sight velocities $v_{y}$ for BS1, the distances $y$ and line of sight velocities $v_{y}$ for CS1.}
\label{figbs1cs1}
\end{center}
\end{figure}

Consider the set of orbits shown in Figure \ref{figlogseries}. These were integrated in a logarithmic halo given by \citep{Binney:2008p5426}:
\begin{equation} \label{eq:logpot}
 \Phi_{halo} = \frac{1}{2}V_{0} ^2 \ln\left(R_{c}^2+R^2 + \frac{z^2}{q_{\phi}^2}\right) \, 
\end{equation}
\noindent in  cylindrical coordinates, where $q_{\phi}$ is the potential flattening, $V_{0}$ is the circular velocity, and $R_{c}$ is the core radius. For this initial test, we neglect the contribution of the bulge and the disk to the potential. The orbital and potential parameters of each of these are listed in Table \ref{tablelogseries}. A logarithmic halo has the advantage of having only three parameters, thereby minimizing possible degeneracies between the different models. Using only the projected positions of these orbits, we find that the flattening $q_{\phi}$ and initial line of sight distance of the orbit from the center of the galaxy $y_{0}$ are easily constrained as shown in Figures  \ref{figqplots} and  \ref{figyplots}, whereas the circular velocity is degenerate and cannot be constrained with only positional information (Figure \ref{figvcplots}). It is to be noted that there may be a sign discrepancy in the distance estimate, as orbits with either value of $y_{0}$ will be identical in projection. Therefore, we restrict the fitting routine to positive $y_{0}$ space and only the magnitude of the progenitor distance is recovered. The orbits in this set are fairly long and have more than two turning points. However, several observed streams are much shorter. In order to analyze the effect of the length of an orbit and more importantly, the number of turning points of the orbit, on the estimate of the flattening, we fit shorter versions of the orbit AS1 (Figure \ref{figbs1cs1}). Not surprisingly, having fewer turning points on an orbit causes a spread in the estimated value of $q_{\phi}$. Adding more information to the fitting (line of sight velocities for BS1 and line of sight velocities and distances $y$ for CS1) results in much more accurate estimations of $q_{\phi}$ as shown in the bottom panels of Figure \ref{figbs1cs1}. As for the distance to the progenitor, $y_{0}$, it is strongly constrained (with a sign discrepancy) for BS1 even only with the projected positions and for CS1, if kinematic information is provided (but not with only the projection), making it the most easily constrainable parameter. 

\begin{table*}
\caption{Series of Logarithmic Orbits: The following are the potential and orbital parameters of the orbits shown in Figure \ref{figlogseries}. $q_{\phi}, R_{c}$ and $V_{0}$ are the potential flattening, core radius and circular velocity respectively.  $x_{0}, y_{0}, z_{0}$ are the initial positions and $v_{x_{0}}, v_{y_{0}}, v_ {z_{0}}$ the initial velocities used.} 
\begin{center}
\begin{tabular}{| l | l l l | l l l l l l |}
\hline
Name & $q_{\phi}$ & $R_{c}$  & $V_{0} $ & $x_{0}$ & $y_{0}$ & $z_{0}$ & $v_{x_{0}}$ & $v_{y_{0}} $ & $v_{z_{0}}$\\ 
             &           & ($\kpc$)& ($\kms$) & ($\kpc$) & ($\kpc$) & ($\kpc$) & ($\kms$) & ($\kms$) & ($\kms$) \\ \hline
 AS1  & 1.0 & 2.0 & 220 & 30.0 & 30.0 & 30.0 & 250.0 & 10.0 & 10.8\\
 AS2  & 1.0 & 3.0 & 180 & 30.0 & 30.0 & 30.0 & 200.0 & 10.0 & 10.8\\
 AS3  & 1.0 & 5.0 & 200 & 20.0 & 10.0 & 30.0 & -230.0 & 40.0 & 20.0\\
 \hline
 AO1 & 0.72 & 5.0 & 200 & 20.0 & 10.0 & 30.0 & -230.0 & 40.0 & 20.0\\
 AO2 & 0.8 & 1.0 & 240 & 20.0 & 20.0 & 20.0 & 220.0 & 200.0 & 180.0\\
 AO3 & 0.8 & 3.0 & 200 & 20.0 & 20.0 & 10.0 & 20.0 & 200.0 & 180.0\\
 \hline
 AP1 & 1.3 & 3.0 & 200 & 20.0 & 20.0 & 10.0 & 20.0 & 200.0 & 180.0\\ 
 AP2 & 1.2 & 3.0 & 200 & 20.0 & 20.0 & 10.0 & 120.0 & 50.0 & 180.0\\ 
 AP3 & 1.4 & 3.0 & 200 & 40.0 & 30.0 & 30.0 & 50.0 & 50.0 & 150.0\\ 
\hline
\end{tabular}
\end{center}
\label{tablelogseries}
\end{table*}%

\begin{figure*}
\begin{center}
\includegraphics[width = 150mm, height = 100mm]{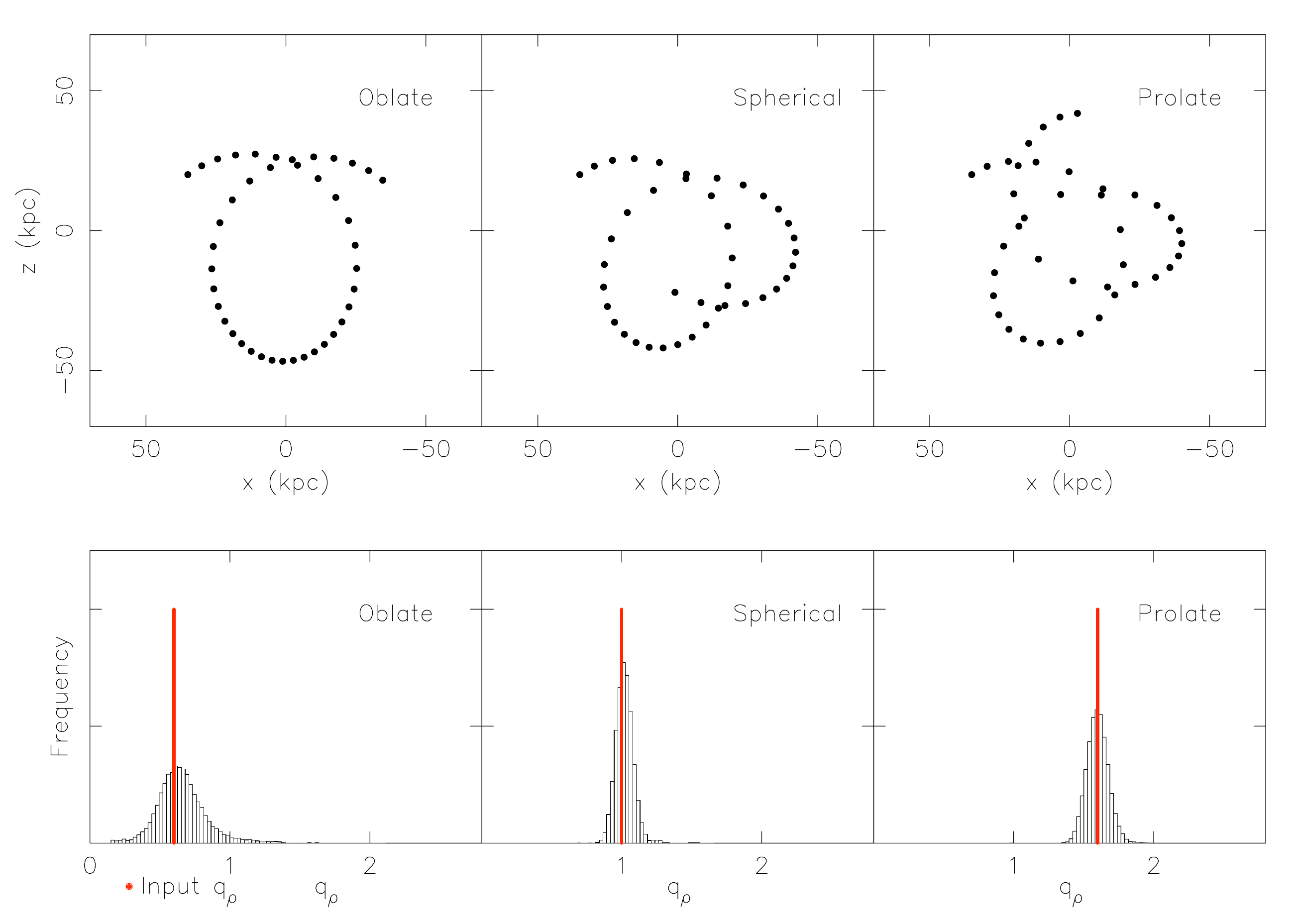}
\caption{Estimation of $q_{\rho}$ for the orbits shown in the top panel. The host galaxy models in these examples are a one-component ellipsoidal halo (Equation~\ref{eq:sphpot}). The input value of $q_{\rho}$ in each case is marked in red. These distributions are drawn from 2,000,000 steps of the coldest MCMC chain.}
\label{figtestsphorbits}
\end{center}
\end{figure*}

\begin{figure*}
\begin{center}
 \includegraphics[width = 150mm, height = 50mm]{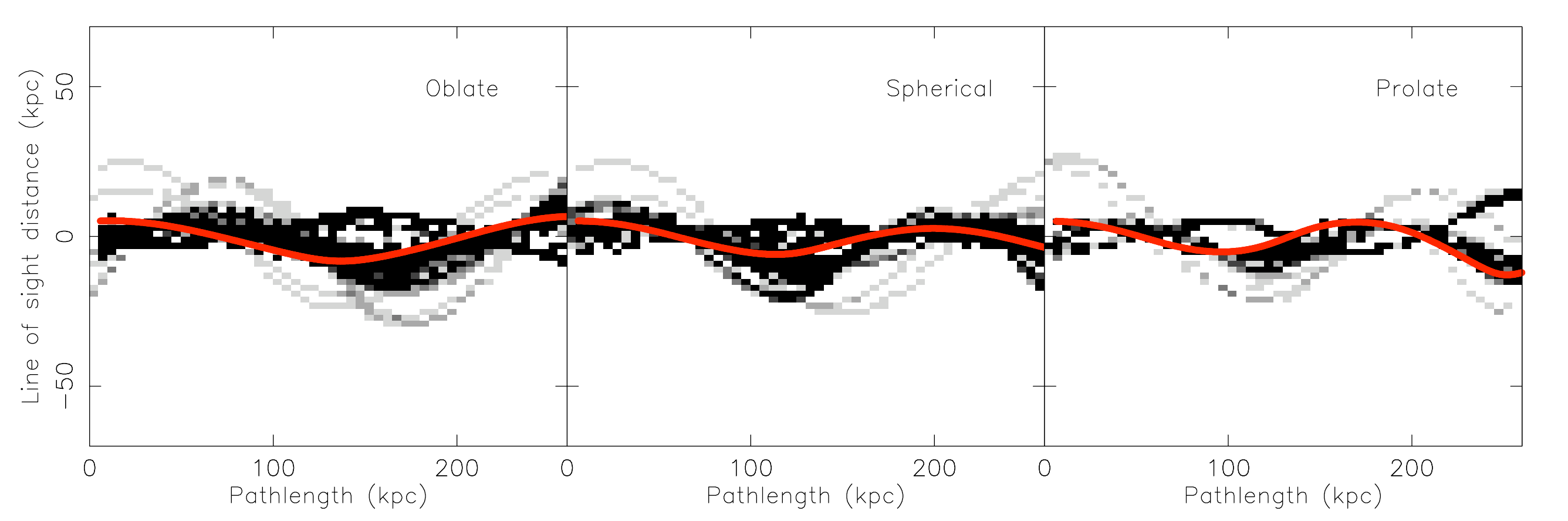}
\caption{Recovery of the three dimensional structure of orbits with only their projections on the $XZ$ plane. The grayscale image shows the density of the line of sight distances along the trial orbits considered in the coldest MCMC chain used in the fitting of the orbits shown in Figure \ref{figtestsphorbits}. The red line marks the actual distance along the input orbits. This figure shows that coarse estimates of the distances along the orbit can be obtained.}
\label{figytestgray}
\end{center}
\end{figure*}

\subsection{Orbits in a Spheroidal Halo}

A more general halo is described by a double power law density distribution \citep{Dehnen:1998p61}
\begin{equation} \label{eq:sphpot}
\rho_{s} = \rho_{0}\left( \frac{s}{r_{0}}\right)^{-\gamma} \left( 1 + \frac{s}{r_{0}}\right)^{\gamma - \beta} e^{-s^2/r_{t}^2} \, ,
\end{equation}
\noindent where $s$ is the ellipsoidal coordinate
\begin{equation}
s \equiv  \left( R^2 + z^2/q_{\rho}^2\right)^{1/2} \, .
\end{equation}
The model parameters that are unknown are the central density $\rho_{0}$, the inner and outer slopes $\gamma$, $\beta$, the scale radius $r_{0}$ and the flattening in density $q_{\rho}$. We calculate the potential due to this distribution by multipole expansion using an algorithm similar to `GalPot' \citep{Dehnen:1998p61, Binney:2008p5426}. The truncation radius $r_{t}$ is fixed at $1000 \kpc$, but its precise value does not affect orbits at the radial distances under consideration. We find that the inner slope $\gamma$ cannot be constrained by orbits that are far away from the center of the galaxy and that small variations in $\gamma$ do not affect the fits. Hence, we adopt a fixed value of $\gamma = 1$, equivalent to the central power-law slope of a NFW profile (\citealt{Navarro:1996p3805}), for the orbital fitting and set the remaining parameters to be free. Figure \ref{figtestsphorbits} shows the distributions in the flattening $q_{\rho}$ obtained by fitting the projections of three orbits in this power-law spheroidal halo. Even though the distributions peak at the right values of $q_{\rho}$, they are much more spread out than the previous distributions in $q_{\phi}$. This is to be expected, as orbits respond directly to the potential, and these are rounder than their corresponding mass distributions. For a logarithmic potential, at distances much larger than the core radius, the relation between $q_{\phi}$ and $q_{\rho}$ can be approximated as \citep{Binney:2008p5426}: 
\begin{equation} \label{eq:qphirho}
 1 - q_{\phi} \approx \frac{1}{3} \left(1 - q_{\rho} \right) \, .
\end{equation}
The fitting mechanism also turns out to be extremely useful in approximating the line of sight distances along the orbit as revealed by a grayscale plot of the distances along the various trial orbits on the coldest MCMC chain (Figure \ref{figytestgray}).

\section{Testing with streams in a spheroidal potential}
\label{streamfitting}

\begin{figure}
\begin{center}
 \includegraphics[width = 80mm, height = 80mm]{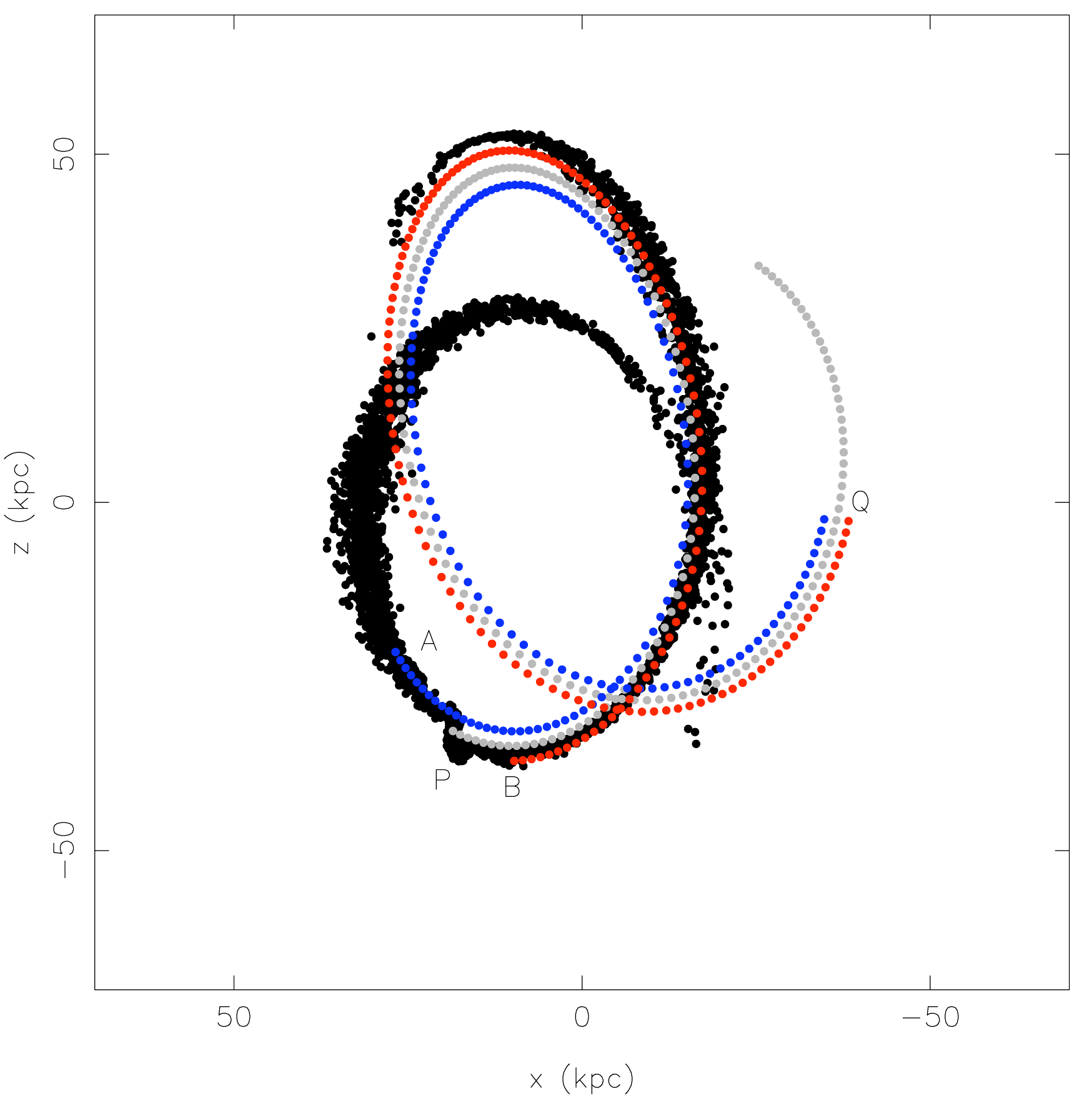}
\caption{Correction Mechanism: Calculating a pair of corrected points on the leading and trailing tails corresponding to a point on the past orbit of the progenitor. The black dots are particles from an N-body simulation of the stream. The grey dotted curve shows the backward integrated orbit of the progenitor, starting from its current position at $P$. $Q$ is an arbitrary point on this orbit. The blue and red dotted curves are the orbits of stars which escape the progenitor system at $Q$ from the inner and outer escape points respectively. The end points of these orbits lie on the leading and trailing arms of the stream, marked as $A$ and $B$. The velocity components at $Q$ of the blue and red orbits are the same as that of the grey orbit (with the opposite sign). This process is repeated for several points along the grey curve to obtain a set of corrected points like $A$ and $B$ that demarcate the stream path (see Figure \ref{figcorrection}).}
\label{figexpcorr}
\end{center}
\end{figure}

\begin{figure*}
\begin{center}
 \includegraphics[width = 180mm, height = 90mm]{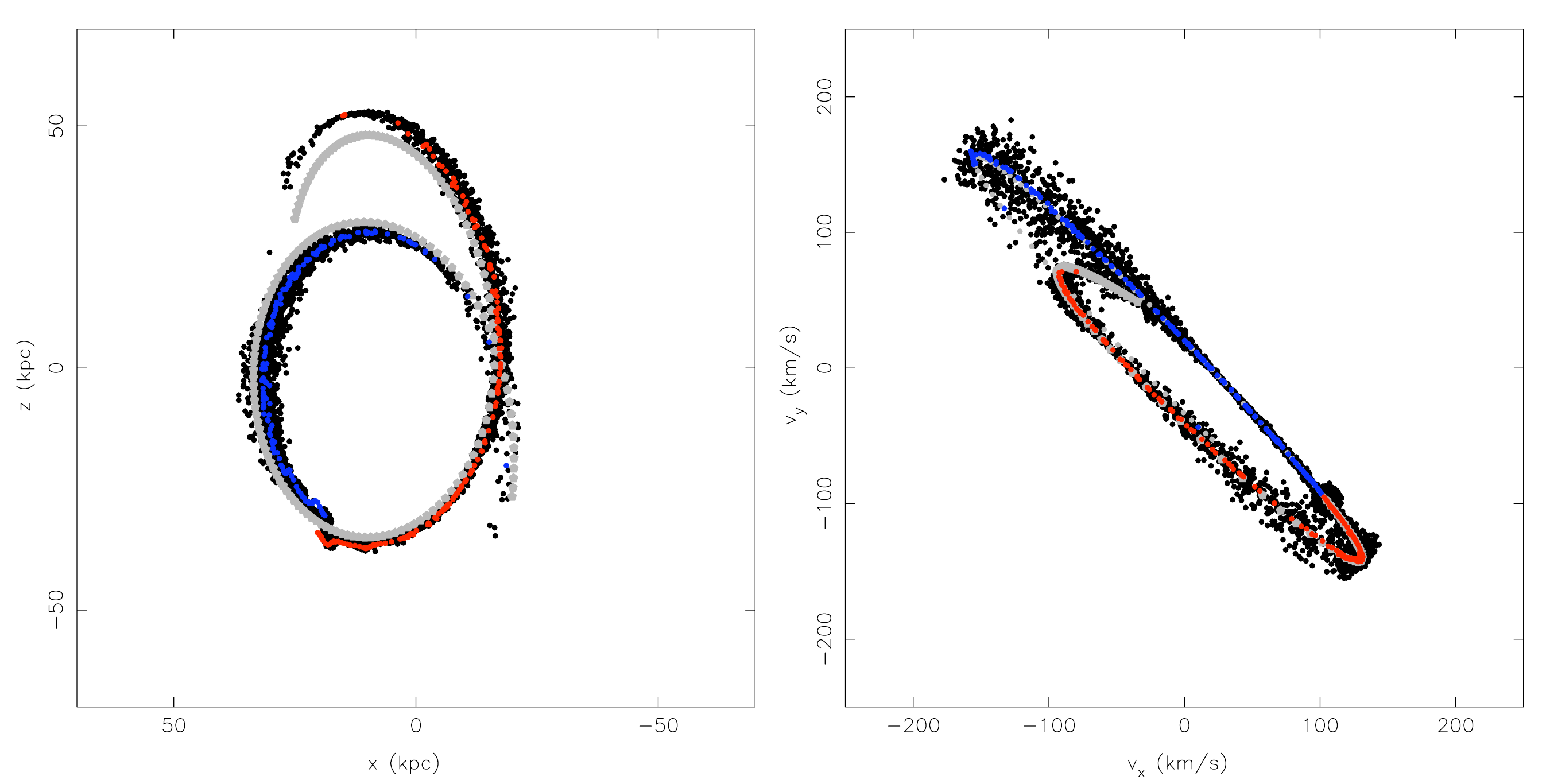}
\caption{Correcting for tidal tails: The left and right panels show stream B in the $XZ$ and $V_{x}V_{z}$ planes, respectively. The black dots are particles from an N-body simulation of the stream. The grey dotted curve is the orbit of the progenitor, the remnant of which is the concentrated sphere. The blue and red dots are the corrected points for the leading and trailing arms, respectively. The very close correspondence between the computationally-expensive N-body simulation and the locus of the ``corrected stream'' points is evident. }
\label{figcorrection}
\end{center}
\end{figure*}

\begin{figure*}
\begin{center}
\includegraphics[width = 120mm, height = 120mm]{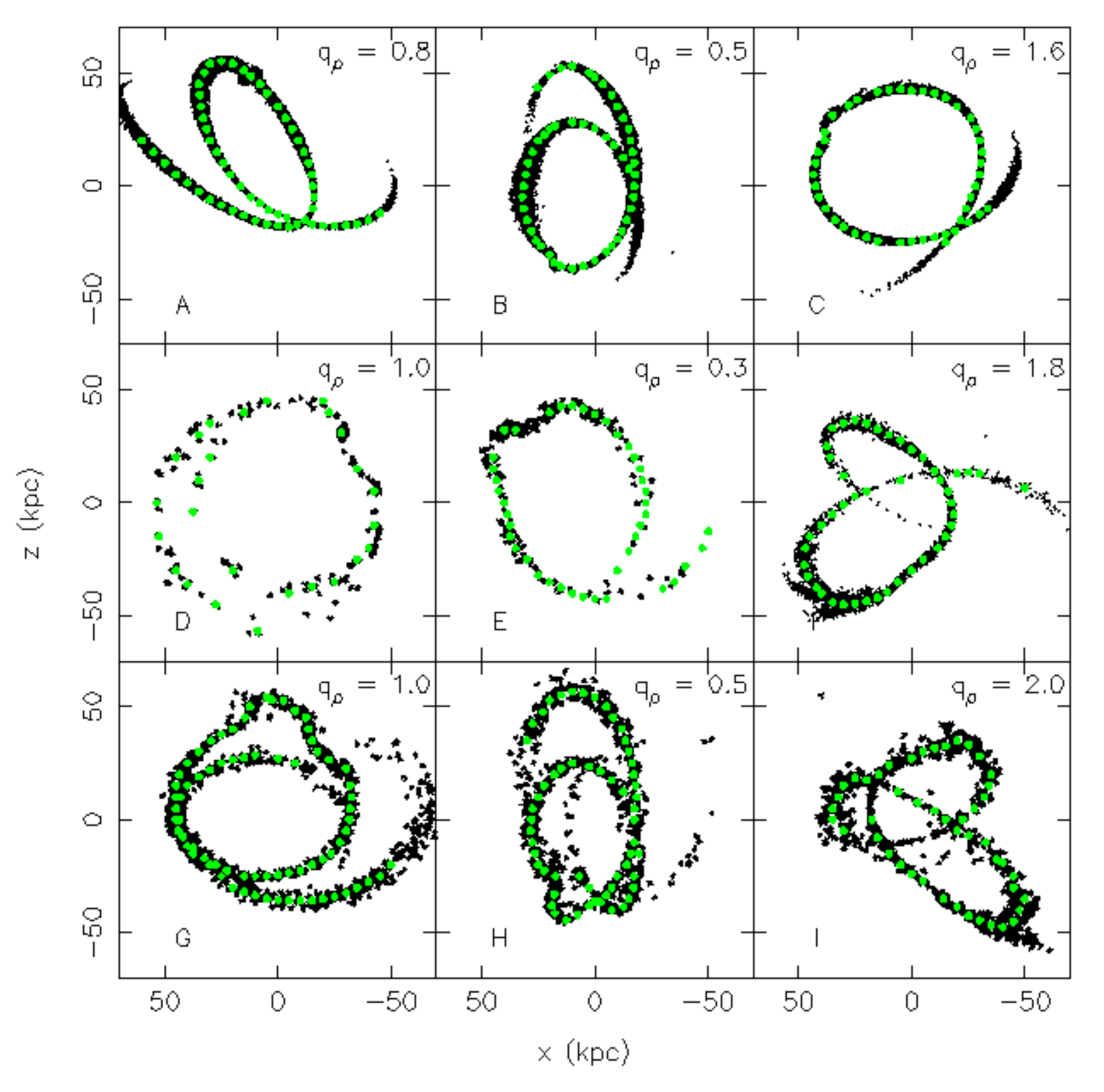}
\caption{The projection of test streams generated by N-body integrations using \emph{gyrfalc(ON)} in a realistic multi-component galaxy model with bulge, disk, thick disk, ISM and halo components. The $q_{\rho}$ value indicated on each panel is the flattening of the halo used in each simulation. The streams in the top, middle and bottom panels were generated from satellites with initial masses of $5.0 \times 10^{6} \msun$, $5.0 \times 10^{7} \msun$, and $1.0 \times 10^{8} \msun$ respectively. The green dots are the points taken along the streams to fit trial streams to.}
\label{figstreams}
\end{center}
\end{figure*}

\begin{figure*}
\begin{center}
\includegraphics[width = 120mm, height = 120mm]{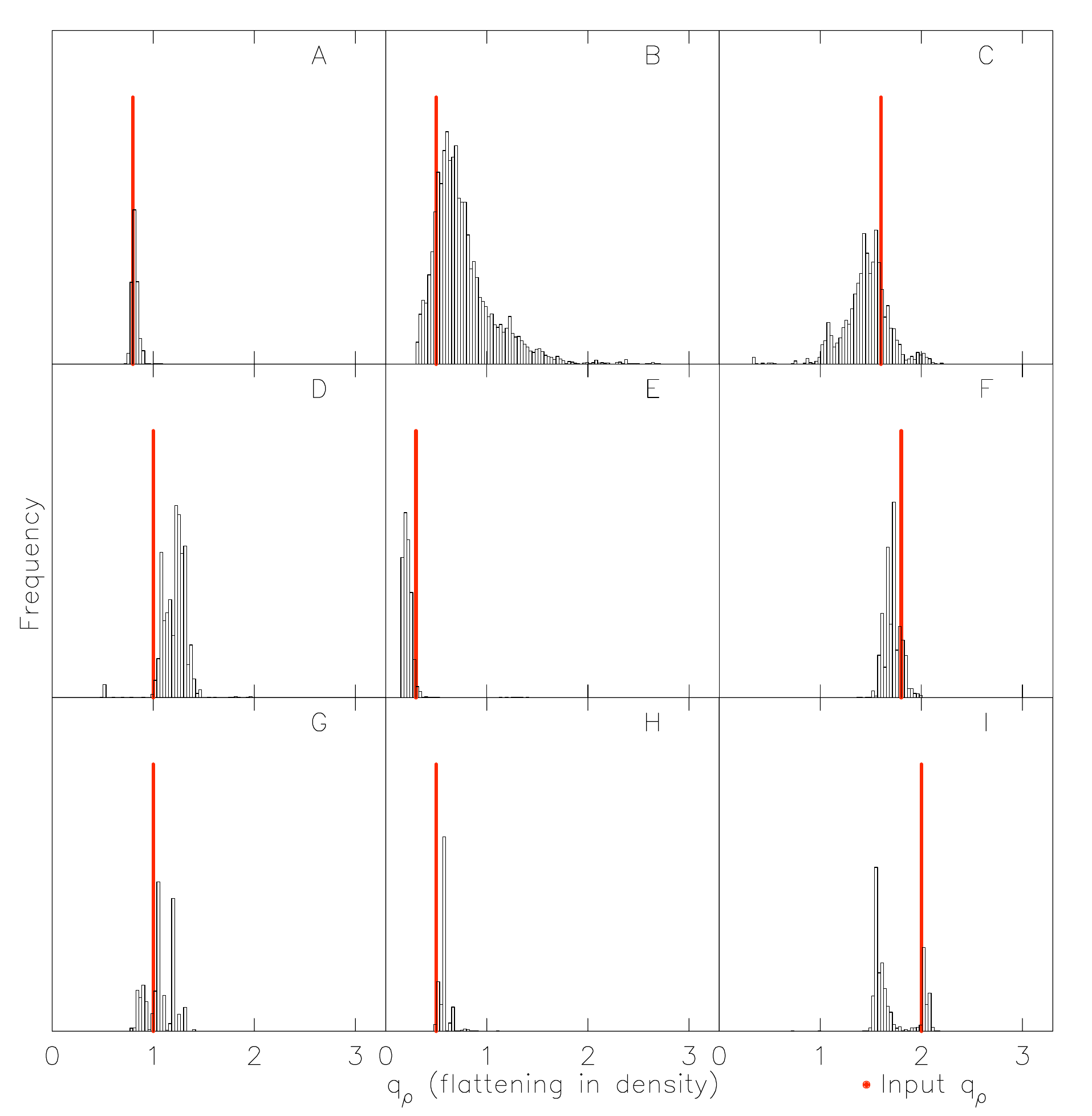}
\caption{Estimation of the density flattening $q_{\rho}$ with only projections of streams. Shown here are the distributions of $q_{\rho}$ obtained by fitting the projection of streams shown in Figure \ref{figstreams}. The number of progenitor orbits attempted vary from $500,000$ to $1,000,000$ from stream to stream. The red lines indicate the input $q_{\rho}$ for each. We find that the accuracy of estimation varies from stream to stream. In all the cases, the distributions indicate the right shape of the halo (oblate, spherical or prolate). For fitting streams B and E, the inner power slope $\gamma$ was kept fixed at 1.0 though its true value is 1.28. This does not introduce any significant error on the estimate of the density flattening. For fitting stream H, we set $\gamma$ to be a free parameter and this does not affect the estimate on $q_{\rho}$ either. For stream I, there are two peaks in the distribution, the actual value of $q_{\rho}$ being at the smaller peak. However, the streams generated at the smaller peak resemble the test stream in finer details, such as its bifurcations, which were not included in the fitting data.}
\label{figqstreams}
\end{center}
\end{figure*}

\begin{figure*}
\begin{center}
\includegraphics[width = 120mm, height = 120mm]{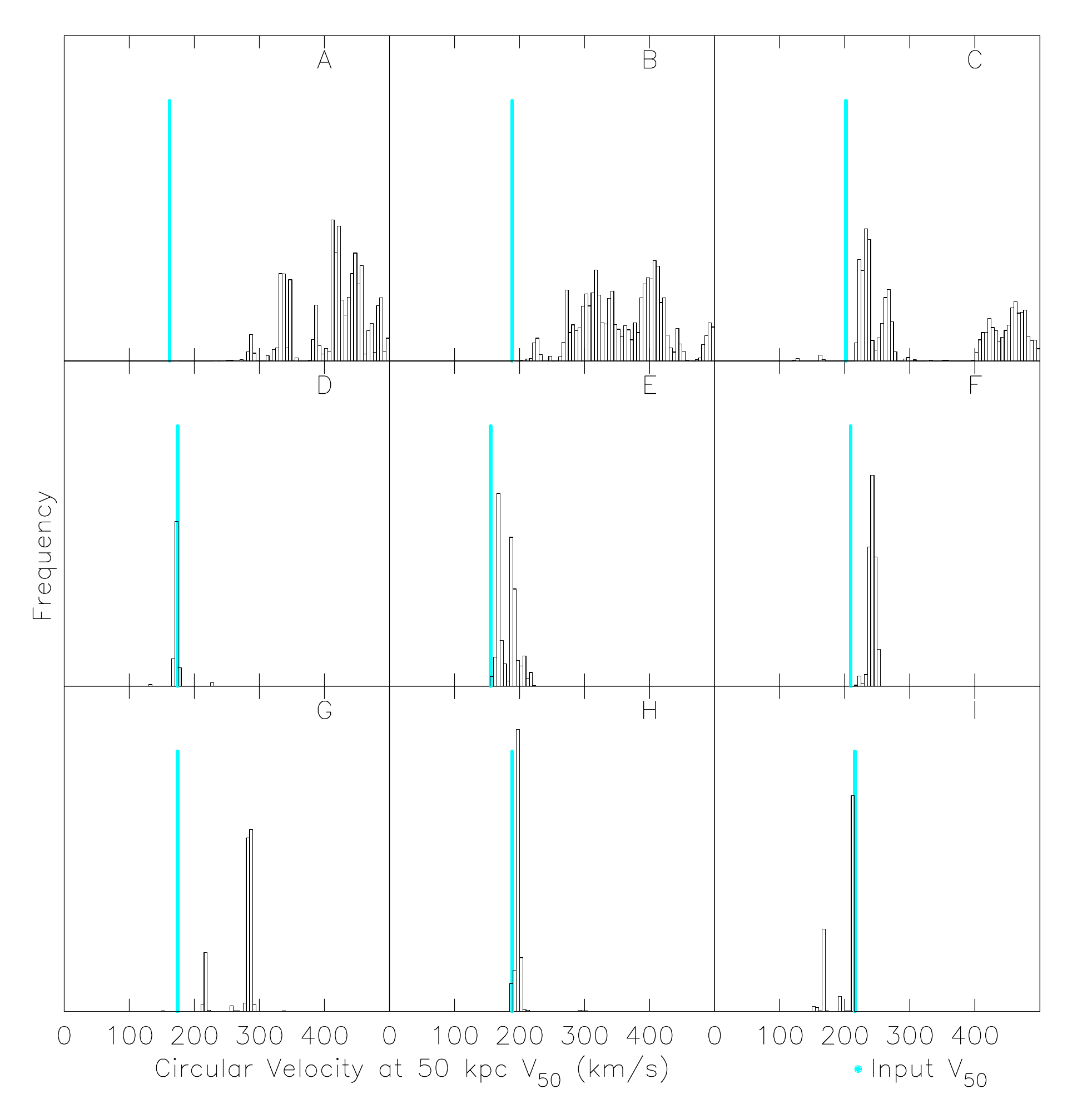}
\caption{Estimating the mass of the halo with only projections of streams. We use the circular velocity at $50\kpc$, $V_{50}$, as an indirect mass parameter. The value of $V_{50}$ is calculated using the potential parameters at each step of the coldest MCMC for the fitting of the streams shown in Figure \ref{figstreams}. These are the distributions in $V_{50}$ thus obtained for these streams. The cyan lines indicate the input $V_{50}$ for each. It is seen that it is impossible to estimate $V_{50}$ and the mass of the halo with only positional information. The algorithm finds highly likely solutions at different values of mass of the halo. This is because several streams (like orbits) exist that are identical in projection, but different in velocity space, for different values of the halo mass. As a result, the algorithm, in many cases, does not even reach the correct input $V_{50}$ value.}
\label{figV50streams}
\end{center}
\end{figure*}

\begin{figure*}
\begin{center}
\includegraphics[width = 120mm, height = 120mm]{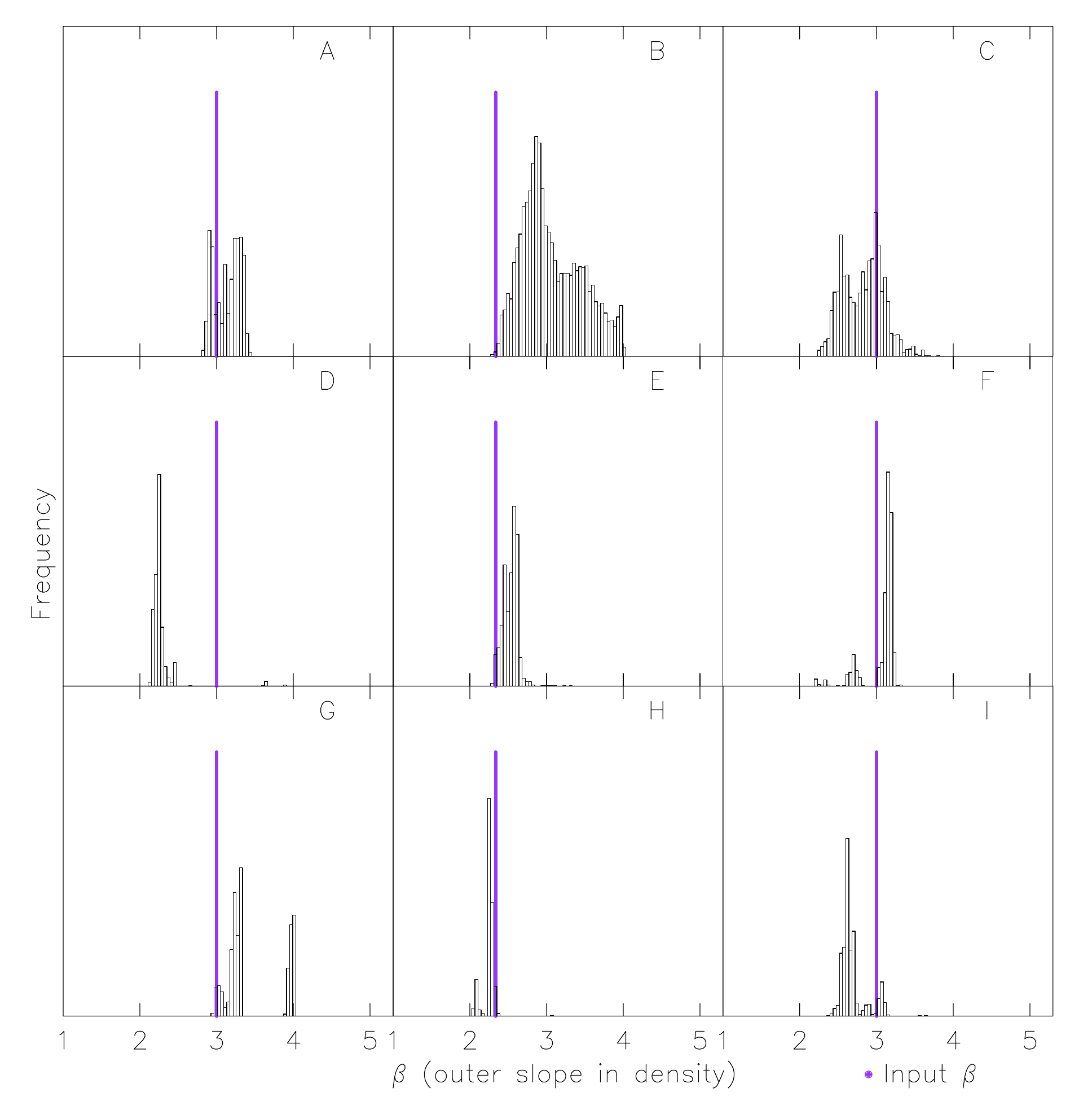}
\caption{Estimating the outer power law $\beta$ with only projections of streams. Shown here are the distributions of $\beta$ obtained by fitting the projection of streams shown in Figure \ref{figstreams}. The violet lines indicate the input $\beta$ for each. This shows that $\beta$ cannot be constrained by the 2-dimensional positional information alone.}
\label{figbetastreams}
\end{center}
\end{figure*}

\begin{figure*}
\begin{center}
 \includegraphics[width = 150mm, height = 50mm]{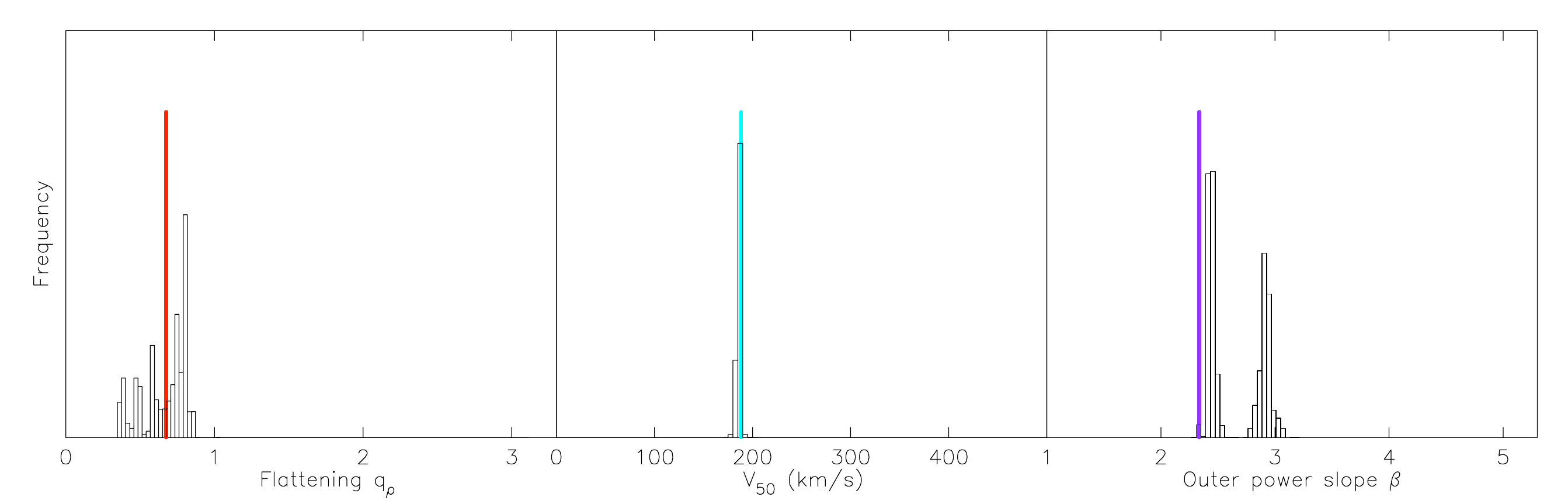}
\caption{Effect on the estimates of parameters of adding the rotational velocity curve of the host galaxy to the projection of a stream. The panels from left to right show the distributions in the density flattening $q_{\rho}$, circular velocity at 50\kpc \, $V_{50}$, and outer power law $\beta$ respectively, for stream B, when the inner circular velocity curve (which extends up to 30\kpc\ for this case) is also provided in addition to the projected positions. The true values of each of these are marked in red, cyan and violet respectively. It is seen that the provision of the rotational velocity curve accurately constrains the $V_{50}$ value as expected. This, in turn, helps constrain other parameters of the model, such as $\beta$.}
\label{figvcfits}
\end{center}
\end{figure*}

\begin{figure*}
\begin{center}
\includegraphics[width = 150mm, height = 100mm]{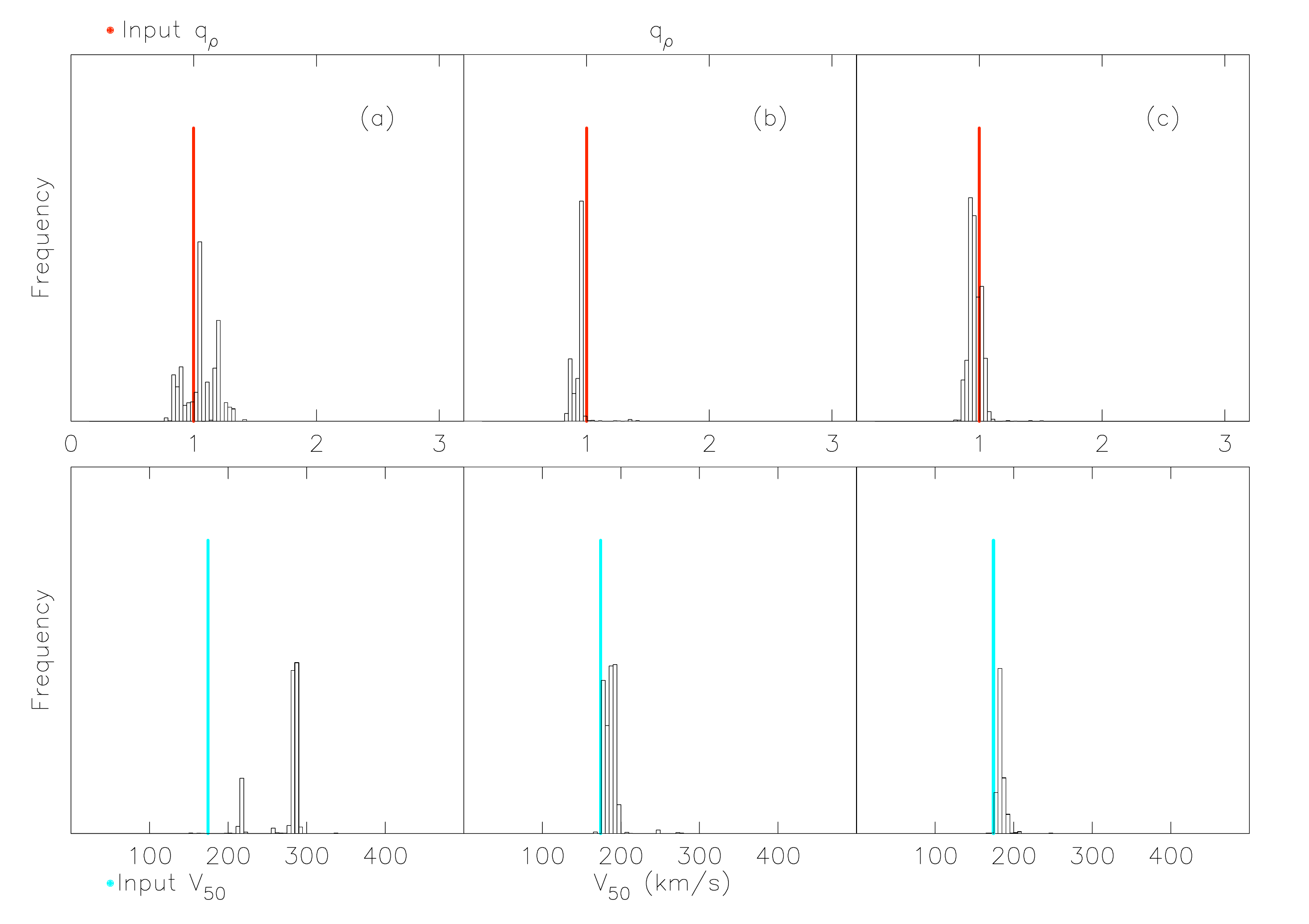}
\caption{Effects on the estimates of $q_{\rho}$ and $V_{50}$ when l.o.s velocities are added to the projection of the stream. The top panels show the distributions in $q_{\rho}$ and the bottom panels the corresponding distributions in $V_{50}$ for stream G, that was generated in a perfectly spherical potential. The true values of $q_{\rho}$ and $V_{50}$ of the model are marked in red and cyan respectively. In (a) only the projection of the stream on the sky is used, and a reasonably good estimate of the flattening obtained, whereas the estimate of $V_{50}$ is poor. Adding the line of sight velocity $v_{y_{0}}$ of the progenitor to the projection of the stream (but no rotational curve) greatly improves the estimates of $V_{50}$ as well as $q_{\rho}$ (b). Adding to this five l.o.s velocity data points for each tail further improves the estimation of these parameters (c).}
\label{fighevsph}
\end{center}
\end{figure*}

\begin{figure*}
\begin{center}
\includegraphics[width = 120mm, height = 120mm]{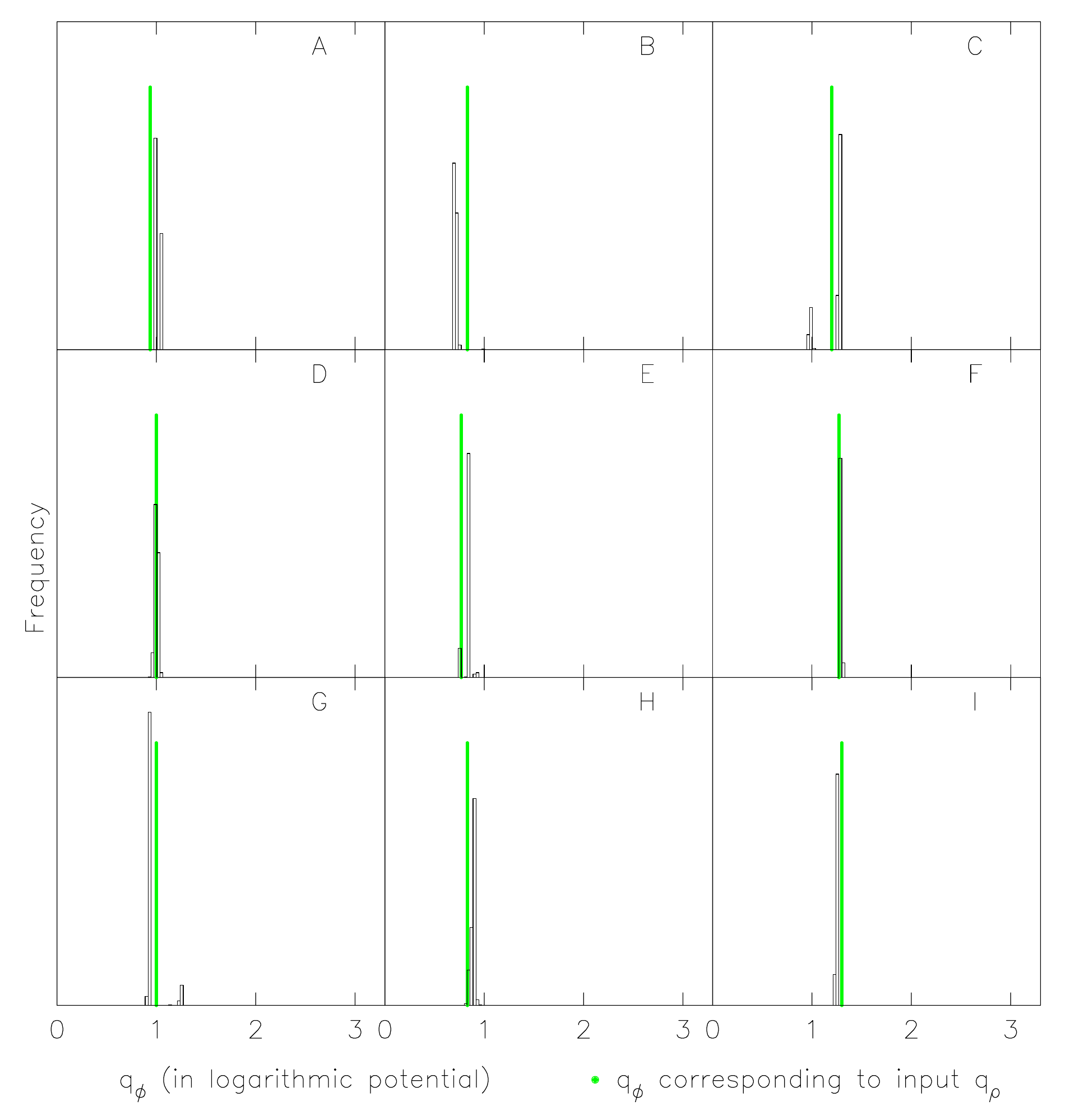}
\caption{Distributions of $q_{\phi}$ obtained when the test streams (generated in a multicomponent host potential with a double spheroidal halo) are fit in a logarithmic potential, without any stellar components. The green lines indicate the flattening in potential $q_{\phi}$ corresponding to the input flattening in density $q_{\rho}$ of the host halo calculated using equation \ref{eq:qphirho}. It is seen that by using a logarithmic halo to fit streams that were generated in a double power law density model, the flattening in the potential can still be recovered.}
\label{figqlog}
\end{center}
\end{figure*}

\begin{figure*}
\begin{center}
\includegraphics[width = 170mm, height = 90mm]{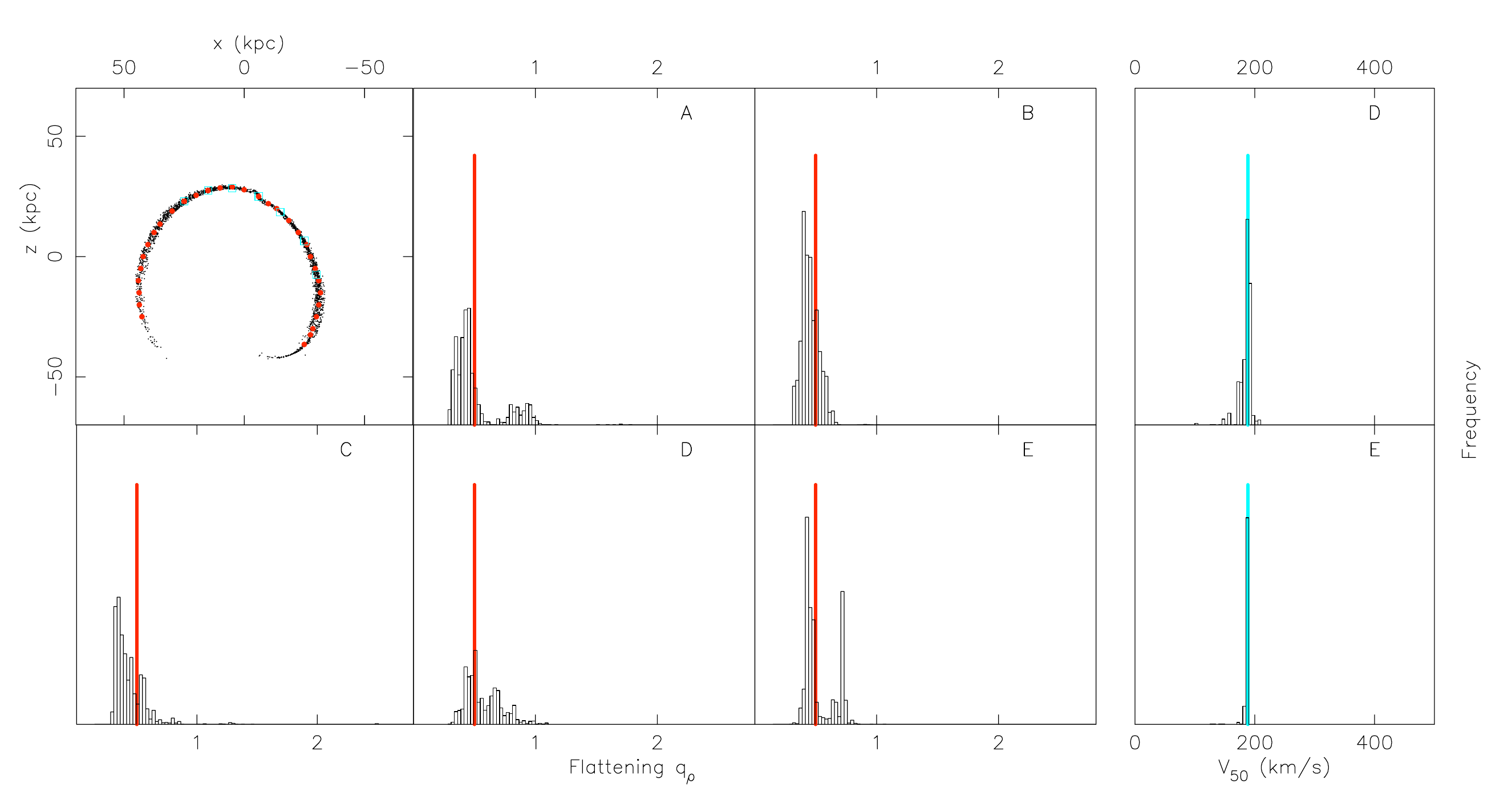}
\caption{Estimation of $q_{\rho}$ and $V_{50}$ for a short stream with only one prominent turning point. The red line shows the input value of $q_{\rho}$. The distributions are drawn from 500,000 steps of the coldest MCMC chain. Case A: only projected positions, red dots show the points used in the fitting. Case B: projected positions and l.o.s velocities at cyan squares. Case C: projected positions and distance to the progenitor. Case D: projected positions, distance to the progenitor and l.o.s velocities at cyan squares. Case E: Same as Case D but with the rotational velocity curve given. The rightmost panels show the estimation of the circular velocity at 50 kpc for case D and E, the cyan lines indicating its true value. It is not possible to constrain $V_{50}$ without any velocity information, but if l.o.s velocities are provided (case D), it can be estimated even with a short stream like the one above. It is not surprising that $V_{50}$ is very well constrained in case E as the circular velocities up to 30~kpc are given.}
\label{figparabola}
\end{center}
\end{figure*}

\begin{figure*}
\begin{center}
\includegraphics[width = 170mm, height =90mm]{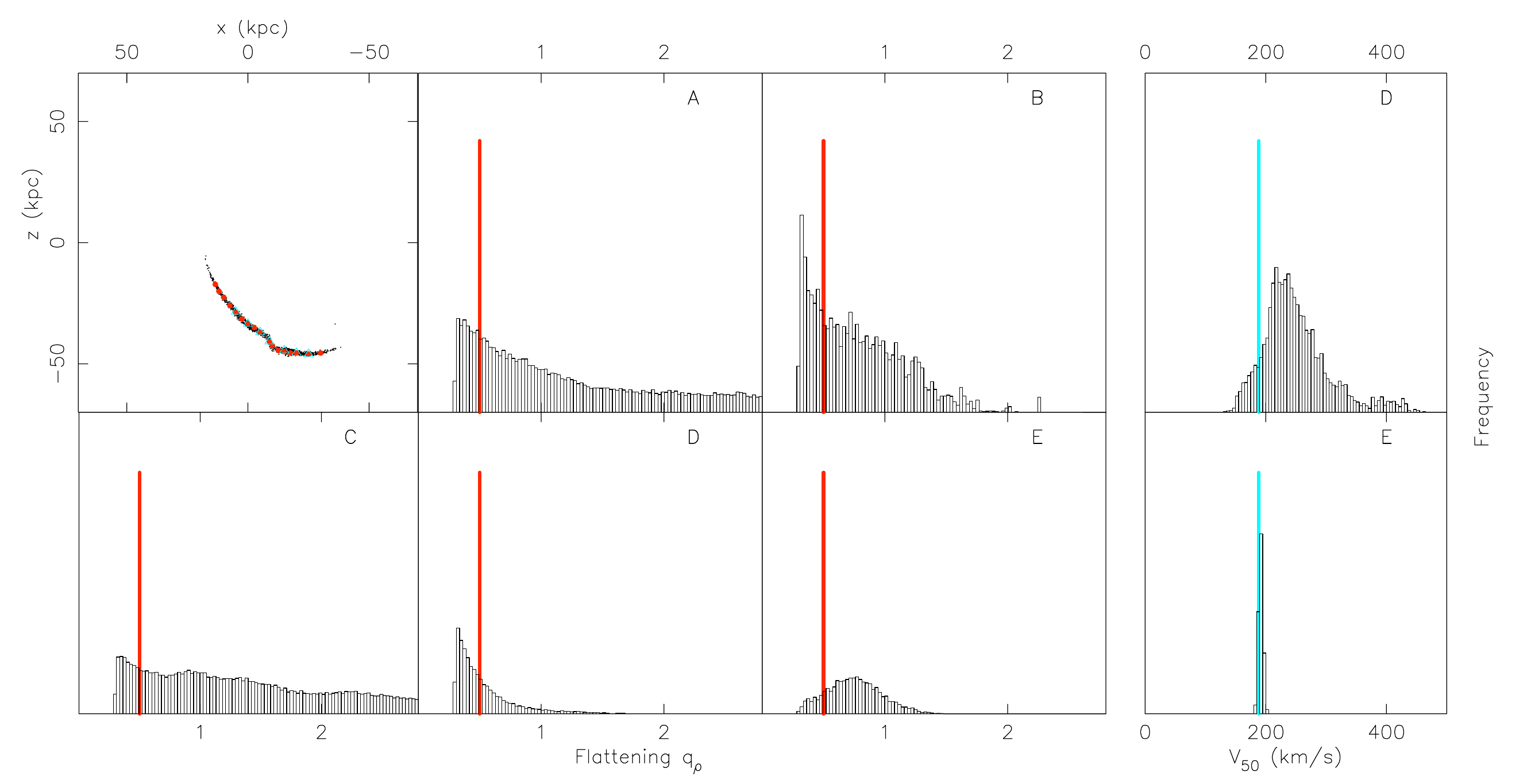}
\caption{As Fig.~\ref{figparabola}, but for a Palomar 5-like stream with no turning point.}
\label{figstrip}
\end{center}
\end{figure*}

Having demonstrated the power of the technique in constraining the parameters of a density distribution by fitting only the projected positions of orbits, we test its ability to robustly estimate the same by fitting streams which, as mentioned in \S\ref{Introduction}, do not delineate the orbit of the progenitor or any other exact orbit in the potential. A consistent and fast mechanism is required to derive the positions of stream stars from the progenitor's orbit, without using N-body integration. The stars which make up the stream are those which were tidally ripped from the satellite during pericenter passages. Based on this, it is possible to formulate a simple correction mechanism that maps a given progenitor orbit to the coordinates of its tidal tails. 

Consider the stream shown in Figure \ref{figexpcorr}, whose projection in the $V_{x}V_{z}$ plane is shown in the right panel of Figure \ref{figcorrection} (it is the same as stream B in Figure \ref{figstreams}, which was integrated in an oblate halo). The grey dotted curve is the orbit of the progenitor, the remnant of which is the concentrated sphere. The progenitor's orbit corresponds to a set of orbital parameters (the progenitor's current position and velocity components) and potential parameters. The stars that lie on the tidal tails escaped the progenitor at earlier times, from two regions of the satellite: one that is closest to the center of the host galaxy and one that is farthest from it. We refer to these points as the inner and outer escape points respectively, which approximate the inner and outer Lagrange points in a restricted three-body problem. Therefore, the trajectory of a stream star can be integrated with its initial position offset from a certain point in the progenitor's past orbit by a certain distance $r_{cutoff}$, offset outwards for trailing tail stars and inwards for leading tail stars. This is shown diagrammatically in Figure \ref{figexpcorr}. Starting at the current position and velocity of the progenitor $P$, its orbit is integrated backwards in time, marked by the grey dots. For any point on its backward orbit, $Q$, at time $t_{Q}$, the inner and outer escape points are approximated as $r_{Q} - r_{cutoff}$ and $r_{Q} + r_{cutoff}$, indicated respectively by the blue and red dots at $Q$. These provide the initial positions for the orbits of stars that escape the satellite at $Q$. For their initial velocity components, we use the velocity of the progenitor orbit at $Q$. With these initial phase space coordinates, we integrate forward for the same amount of time $t_{Q}$. These orbits are indicated by the blue and red dotted curves, with their final points on the leading and trailing tails at $A$ and $B$ respectively. Repeating this process for several points on the backward integrated orbit (say, for every 50~Myr) yields a set of points which lie closely on the tidal tails of the stream (red and blue dots in Figure \ref{figcorrection}), which we shall refer to as the corrected points for a given set of parameters. 

We find empirically that the offset radius for a point $Q$ at a radial distance $r$ can be calculated approximately as $2.88$ times the theoretical Jacobi radius:
\begin{equation} \label{eq:rcutoff}
r_{cutoff} = 2.88 \times \left(\frac{m_{sat}}{3\s{M}(r)}\right)^{1/3}r\, ,
\end{equation}
where $m_{sat}$ is the mass of the satellite and $\s{M}(r)$ is calculated from the circular velocity $V_{c}$ of the host galaxy at $r$ as $\s{M}(r) =  rV_{c}^{2}(r)/G$, where $G$ is the gravitational constant. For a perfectly spherical potential, $\s{M}(r)$ is equivalent to the mass contained within radius $r$. For non-spherical potentials, it is only a crude approximation to the total mass inside $r$, but as we show below, the correction we obtain using $r_{cutoff}$ calculated in this manner is sufficiently accurate for our purposes.

We find that by fitting the stream data with these corrected points it is indeed possible to recover the parameters of the potential as well as the orbit. It is interesting to note that through this correction mechanism, one has more information on the progenitor's past orbit than one would have with only the local orbit of the progenitor (i.e. the red and blue dotted curves in Figure \ref{figcorrection} contain more information than the grey curve does). In this sense, the tidal stream retains some memory of its infall. It is important to note that although we do not correct for velocities at the escape points, the velocities of the corrected points also lie along the velocities of the tidal tails as seen in Figure \ref{figcorrection}, i.e. the correction is valid in all of phase space. This is true of all the streams we tested, possibly because we use low mass satellites. If a similar correction mechanism is used for extending the present technique to more massive streams, we expect that a corresponding correction in velocity space would be required.

The backward integration time depends on the age of the stream and is usually unknown. It is also assumed in the present discussion that the progenitor is identifiable and its position on the sky is known, but more often than not, the satellite is completely disrupted or obscured and it is impossible to identify the progenitor (for example, the NGC 5907 and the Monoceros streams). The time for backward integration can be found empirically, running different fitting routines with different backward integration times for the correction (it only need be accurate to around 1 Gyr). Similarly, several fitting routines can be tried assuming the unidentified progenitor to be at different positions along the stream. However, in all the tests described in this paper, the progenitor, its initial mass and backward integration time, were taken to be known quantities. 
 
Figure \ref{figstreams} shows nine streams on which this stream-fitting method was tested (note that we are now fitting streams, not orbits as in \S\ref{Testing}). These were generated with \emph{gyrfalc(ON)}, an N-body integrator based on a fast and momentum conserving tree-code  (\citealt{Dehnen:2000p3474, Dehnen:2002p3485}), as implemented in the stellar dynamics toolbox NEMO \citep{Teuben:1995p6016}. For the host galaxy, realistic models were used that comprised a spheroidal halo, a spheroidal bulge, and exponential disks (thick disk, thin disk and ISM) similar to the ones described in \cite{Dehnen:1998p61} for the Milky Way. Except for the flattening of the halo, the model of the host galaxy for each case is one of the two models described in Table \ref{tablemodels}. The halo flattening and the initial conditions of the orbits of the satellites for each stream along with its integration time are listed in Table \ref{tablestreams}. Spherical King models of 100,000 particles were used for the N-body satellite simulations. See Table \ref{tablestreams} for the masses, tidal radii and central potentials of these. We have limited the present study to satellite masses in the range of $5.0 \times 10^{6}$ to $1.0 \times 10^{8} \msun$, as they seemed to be ideal candidates for this technique rather than more massive and heavily disrupted satellites, for which dynamical friction plays a significant role. For the integration with \emph{gyrfalc(ON)}, a Plummer softening kernel, along with a softening length of $0.03\kpc$, and an opening angle tolerance parameter of 0.6 (the default) were used.

We make extensive use of the work of \cite{Dehnen:1998p61} for the calculation of the potential, especially for the bulge and disk components. We assume that these baryonic components of the system are well understood and their contribution to the potential is kept fixed, so that the only parameters to be estimated are those of the dark halo. As the streams are distant, the precise details of the inner mass distribution are not critical and the estimates are not affected by small variations in the baryonic component. The potential due to the trial density distribution of the halo is calculated using multipole expansion. The calculated halo potential and its first derivatives (w.r.t $\ln r$ and $|z/r|$) are stored on a grid in $\ln r$ and $|z/r|$. Adding to these, the potential and accelerations due to the bulge and the disk using GalPot, we have the total potential and acceleration at each of the grid points. The acceleration at any point is obtained by interpolating with a 2-dimensional, fifth-order spline. 

In the following sections, we describe the results of fitting the streams with different kinds of information.

\subsection{With projected positions alone}

In this section, we discuss how well the various parameters of the halo are constrained by fitting only the projection of streams on the sky, as this is the only information that is usually available for very distant systems. Figure \ref{figqstreams} shows the distributions of the density flattening $q_{\rho}$ obtained by fitting the pseudo-streams with corrections to trial orbits described in the previous section. The number of attempted orbits in the fitting routines are in the range of $500,000$ to $1,000,000$, running the fitting algorithm until the chains are well mixed and reasonable fits of the test streams are found. As can be seen, the estimates fall within reasonable ranges of the true values of $q_{\rho}$ (indicated by the red lines), but the accuracy of the estimates vary from case to case. The differences in the accuracies arise from the differences in the streams themselves, as some configurations are more degenerate, in that many parameter choices can reproduce the same observations. However, even in the poor cases, the estimated $q_{\rho}$ still provides a useful indicator of the shape of the halo. For example, for stream I, made in a prolate halo of $q_{\rho} = 2.0$, there are two peaks in the distribution, the true value of $q_{\rho}$ being at the smaller peak. The trial streams that are generated at both the peak values of $q_{\rho}$ resemble the given stream, but streams generated at the smaller peak (with $q_{\rho} = 2.0$) show extra features such as small bifurcations that are similar to the ones seen in stream I, whereas streams generated at the larger peak (with $q_{\rho} \approx 1.6$) do not show these. Thus, the streams at $q_{\rho} = 2.0$ resemble the test stream in its finer details. These smaller features were not included in the fitting as we deemed them difficult to observe, especially in distant systems, but if observed, they could be used to obtain better estimates or distinguish between multiple solutions. Nonetheless, the distribution in $q_{\rho}$ for stream I is limited to the prolate region and the corresponding distribution in the potential flattening $q_{\phi}$ is narrower (approximately, $q_{\rho} = 1.6 \equiv q_{\phi} = 1.2$ and $q_{\rho} = 2.0 \equiv q_{\phi} = 1.33$, using equation \ref{eq:qphirho}). It should be noted that for streams B and E, the true value of the inner slope $\gamma$ of the halo is approximately 1.28, but they were fitted in a model with a fixed $\gamma = 1.0$. This was primarily undertaken to check if the difference in the inner slope affects the estimation of $q_{\rho}$, and it does not, confirming our assumption that these streams are insensitive to the finer details of the inner matter distribution. On the other hand, for fitting stream H, we set the inner power slope $\gamma$ to be a free parameter, with lower and upper bounds at -1 and 2.0. This does not seem to deteriorate the estimate of $q_{\rho}$ as was initially expected, so $\gamma$ can be set to be a free parameter in the application of the technique to real stellar streams. However, $\gamma$ itself is not found to be constrained by fitting the projections of these streams.

Another significant parameter of the halo is its mass, which determines the circular velocities at large radii. In order to see how well our technique estimates this quantity, we plot the distribution of the circular velocity at 50\kpc, $V_{50}$, shown in Figure \ref{figV50streams}, calculated from the potential parameters at each step of the coldest MCMC chain. As seen in the case for logarithmic orbits, these distributions are degenerate and it is not possible to constrain the halo mass with only positional information of a stream. We find that the remaining parameters of the halo, i.e. the outer slope $\beta$, the central density $\rho_{0}$ and the scale radius $r_{0}$ can have different values for the same value of the flattening $q_{\rho}$ to produce streams that look identical in projection. The distributions of these parameters are degenerate (as shown in Figure \ref{figbetastreams} for $\beta$) and the projection of a stream alone cannot constrain them. The combination of these parameters result in different circular velocities, making the distribution of $V_{50}$ degenerate. 

As for the orbital parameters, the line of sight distance $y_{0}$ of the progenitor from the center of the galaxy  is well-constrained (although there may be a sign discrepancy, as was seen in the case of pure orbits), whereas the line of sight velocity or the tangential velocity components of the progenitor cannot be recovered.

\subsection{Adding circular velocities}

For many spiral galaxies, inner rotational velocities are available (for instance from \ion{H}{1} kinematics) or are feasible to measure with current instrumentation, which provides information on the inner mass profile of the galaxy. Figure \ref{figvcfits} shows the effects of adding the rotational velocities (here assumed to extend up to 30\kpc) to the projected positions for stream B (to take a particular example). Comparing the corresponding distributions for the stream in Figures \ref{figV50streams} and \ref{figbetastreams}, we see that there is a marked improvement in the estimate of the outer power slope $\beta$. This is, as explained in the previous section, due to the correlation of $\beta$ and the circular velocity parameter $V_{50}$, the latter being estimated accurately with the provision of the inner circular velocities. The peaks of the distributions in $q_{\rho}$ in Figure \ref{figvcfits} and Figure \ref{figqstreams} are at similar positions, but the former is less smooth as a different proposal step size was used to fit the stream in this case. Though not shown here, the degeneracies in the scale radius $r_{0}$ and the central density $\rho_{0}$ are also removed and we obtain accurate estimates of these parameters. Moreover, distance to the progenitor and its velocities can also be constrained when the rotational curve is added to the fitting data. As seen in the distance parameter, there may be a sign discrepancy in the estimate of the line of sight velocity $v_{y_{0}}$ of the progenitor, but not in the estimates of its tangential velocity components.

\subsection{Adding line of sight velocities}

Using stream G as an example, we illustrate the effects of adding kinematic data to the projected geometry of the stream (but without circular velocities) on the estimates of $q_{\rho}$ and $V_{50}$. Figure \ref{fighevsph} shows the $q_{\rho}$ distributions in each case in the top panels and the $V_{50}$ distributions in the bottom panels. In (a), only the projected positions of stream G on the $XZ$ plane are used in the fitting routine. In addition to these, the line of sight velocity of the progenitor is provided to obtain the distributions for (b). To make the distributions in (c), we added more line of sight velocities (at five points along each tail, within $20\kpc$ from the progenitor). This shows that by providing velocity information it is possible to estimate $V_{50}$ or the mass of the halo. The added information also improves the estimate on $q_{\rho}$. As in the case where we provided the rotation curve, the progenitor's velocity components and its line of sight distance are well constrained. 

\subsection{Fitting in a logarithmic potential}

We also investigated how well the test streams in Figure \ref{figstreams} (simulated within a full multi-component galaxy model) can be fit with a simple axisymmetric logarithmic halo with no stellar components. Figure \ref{figqlog} shows the distribution in $q_{\phi}$ thus obtained. The green lines indicate the potential flattening $q_{\phi}$ that correspond to the input density flattening $q_{\rho}$ values. We find that the estimates are remarkably accurate. This exercise suggests that a logarithmic halo can be used as a good approximation to the more complex input axisymmetric model, at least in order to constrain the flattening of the halo. 

\subsection{Shorter streams}
We have seen in \S\ref{logorbits} that the estimation of parameters is sensitive to the number of turning points of the orbit. To explore this aspect of the problem for streams, we consider stream B at much earlier stages of its infall, where its tidal tails are substantially shorter. In Figure \ref{figparabola}, it is at 5~Gyr into its infall and the tidal arms form a parabolic curve, much like the structure observed to the north west of M31 \citep{Carlberg:2011p6167}. In Figure \ref{figstrip}, it is at 2~Gyr in its evolution and resembles the Palomar 5 stream (\citealt{Odenkirchen:2001p2369}). For both these cases, we consider five possible scenarios of available information. Case A is where we only have the projected positions of the streams on the sky. We see that the distributions of $q_{\rho}$ are much more spread out than for stream B, the degeneracy being much more for the Palomar 5-like stream. For Case B, the line of sight velocities at certain points (red squares) are also provided for the fit. In Case C, we assume that no kinematic data is available, but the distance to the progenitor is known. It is possible to measure this quantity for many nearby systems as well as in our own galaxy with the TRGB method. In Case D, the distance to the progenitor and line of sight (l.o.s) velocities are provided in addition to the projected positions. Case E has the maximum information, where the rotational velocity curve is also available along with all the information of Case D. This case has been included primarily to study the effect of circular velocity information on the estimation of $q_{\rho}$. The distributions shown have been drawn from 500,000 steps of the coldest Markov chain.

Here, only the estimates of $q_{\rho}$ and $V_{50}$ are presented. For the stream in Figure \ref{figparabola} with only one prominent turning point, the distributions of $q_{\rho}$ are limited to the oblate region and peak at approximately the true value of $q_{\rho}$ (red lines). This particular example of this class of streams suggests that the estimates of the flattening are much more accurate when kinematic data are available as in cases B, D and E, although the estimate is good even with only the projection of the stream. The rightmost panels show the distribution of $V_{50}$ for cases D and E (the cyan lines mark the input value). Though not shown, it is not possible to constrain this quantity without any kinematic information, as was seen for the longer streams. However, the technique yields very good constraints on $V_{50}$ when a few velocity data points or the inner rotational curve are added to the spatial projection of the stream, making it possible to estimate the mass and shape of the halo for systems where such information is available. 

The short stream shown in Figure \ref{figstrip} has no noticeable turning point. Consequently, the distributions of $q_{\rho}$ are broad and highly degenerate even when the distance to the progenitor or line of sight velocities are provided. It is only when both the progenitor distance and line of sight velocities along the stream are provided that the distribution peaks at approximately the right input value of the flattening. When the circular velocities are also added to the fitting, then the distribution looks more Gaussian-like and a better estimate of $q_{\rho}$ is obtained. One has to bear in mind that the difference in the estimated and true values of the flattening in the potential is expected to be much less than that of the flattening in the density distribution, since the distribution of the corresponding $q_{\phi}$ is narrower as per equation \ref{eq:qphirho}. As for the mass estimates, even with a stream as short as this, the constraints on $V_{50}$ are good when sufficient information (as in case D) is available. Such short streams are mostly observed in systems very close to us, for which the rotational velocity curve, line of sight velocities and distance to the progenitor are available or can be easily measured. Therefore, they can be used as effective probes of the shapes and masses of nearby halos. 

\section{Discussion}
We have shown that stellar streams found in halos of galaxies can be used to constrain parameters of the mass distribution, even without any kinematic information. The easiest potential parameter to estimate, and perhaps one of the most interesting, is the flattening of the distribution $q_{\rho}$. This is under the simplifying assumption that the halo is axisymmetric. In a future contribution, we will test the technique with triaxial halo models. Not surprisingly, our technique cannot constrain the mass of the halo (measured by the circular velocity at $50\kpc$) without any kinematic information. However, by adding the rotational velocity curve or line of sight velocities of the stream stars, it is possible to estimate the total mass as well. According to the currently-held view of hierarchical galaxy formation, large galaxies are built up by the accretion (and disruption) of smaller satellite galaxies \citep{White:1978p1556}. If this is the case, one can expect to find remnants of the process in many galaxies and these will certainly be revealed by future deep surveys (E-ELT, JWST). By applying this technique to a large sample of galaxies, it would be possible to obtain statistics on the shapes of halos, which is crucial in understanding dark matter and galactic evolution. The method is immediately applicable to individual streams, the results of which have several implications. For one, they would provide a better description of the environment in which other phenomena occur. The properties of the halo can be used in studies involving satellite dynamics, tidal debris, gas distributions in the outer regions of galaxies, lensing and many other areas of interest. 

The method could also provide a consistency check for galaxies where the dark matter distribution or its shape has been measured by one of the techniques mentioned in \S\ref{Introduction}. Tidal streams may also be used in testing alternative theories of gravity. For instance, if a prolate halo is found, it could eliminate MOND as a viable theory (\citealt{Read:2005p5282}). In addition to the halo parameters, the technique also recovers the progenitor orbit, giving us a coarse estimate of the distance to the progenitor, and clues to its location if it is unknown, as well as distances and velocities along the stream. 

The correction mechanism employed is one of the key aspects of the technique. It enables us to sample millions of models of the stream without having to resort to N-body simulations for each case, making the problem computationally tractable. During the developmental stages of this paper, we attempted to fit the tidal tails with orbits alone and this sometimes resulted in wrong estimates or highly degenerate distributions of the flattening. Thus, the correction mechanism plays an important role in the effectiveness of the method and the accuracy of the estimates obtained. For its simplicity, it reproduces the stream for an orbit remarkably well, even recreating the bifurcations and other features seen in the stream simulations. The only drawback is that one requires the mass of the in-falling satellite to calculate the corrected points, an approximate value of which can be obtained from star counts or integrating the light along the stream. 

Our correction mechanism has only been tested with spherical, non-rotating King satellites. It is unclear how well it will work for disk-like or other types of satellites, but it should be possible to obtain a correction mechanism for these based on the same principles. However, this means that the results for an observed stream would depend on the assumptions made about its progenitor. For low mass streams such as the ones discussed in this paper, it is safe to assume that the progenitor is a globular cluster or dwarf spheroidal galaxy. Nevertheless, it would be useful to check the correction mechanism with other models of the satellite to see how sensitive it is to the input satellite properties. 

In light of all these assumptions, the technique in its present form is applicable to cold, low-mass streams ($M  \leq 1.0 \times 10^{8} \msun$), for which dynamical friction can be neglected. Observationally, stellar streams fall into two major categories: the ones that are closer to us, in which the positions and velocities of individual stars can be measured and the ones that are farther away and detected in surface brightness. For the latter, the kinematics are usually not measurable. In rare cases, globular clusters associated with the stream may be observed and their velocities can be used as additional information \citep{Mackey:2010p5132}.  Among the streams that are closer to us, are short structures like the Palomar 5 stream. For these, a large number of radial velocity measurements can be made, and these streams can be used to constrain the parameters of the host halo, despite having no turning points. Another plausible scenario is where two or more streams are observed in the same galaxy. Fitting them simultaneously would provide constraints on the halo, even if the individual streams are relatively short. Multiple streams in a halo provide a possible means of probing the halo shape through different cross sections of the halo, which would also enable us to check for triaxiality of the halo.

A surprising result we found, that can serve as a powerful diagnostic tool in the application of the method to real systems, is that if the streams (generated in a host potential with a double power law halo with a bulge and a disk) are fit in an axisymmetric logarithmic potential, the flattening of the potential can still be well estimated. This is encouraging as it suggests that while extending the method for triaxial models, one could use a triaxial logarithmic potential for preliminary estimates on the axial flattenings, which is much easier to use than triaxial density distributions. In addition to the logarithmic and double power law models described in this paper, we also experimented with non-parametric density models for the halo, in which the matter distribution was defined by the values of densities at different radii. Preliminary tests showed that these parameters are difficult to constrain, but this may yet be improved upon. 

\section{Conclusions}
\label{Discussion and Conclusions}
We have extended the idea of using tidal streams to estimate the shape of the dark halos they reside in. The major challenge was that for most streams, only their projected positions are available. This has been a hindrance in using these streams effectively as there may be streams with similar projections for different profiles. We overcome this difficulty by adopting a statistical approach. We sample various orbits with different orbital and potential parameters using parallel MCMC chains. The output distribution of the flattening parameter that we obtain from these peak at the right value, if the information is sufficient. We first tested the method on orbits in a logarithmic potential and then on streams made with N-body simulations in more realistic galactic potentials (sum of disk, ISM, bulge and spheroidal halo). Another major aspect by which the technique differs from earlier work is that the stream in question is not treated as an orbit, but derived as a correction to the orbit of its progenitor.

We have demonstrated that there are cases in which it is possible to get constraints on the shape of the dark halo using only the projection of a tidal stream on the sky, the accuracy of which varies from stream to stream. The method is sensitive to the number of turning points of a stream and only works when there are at least two turning points. With more turning points, the accuracy of the estimate increases. If the stream is only a parabola or turns out to be degenerate even with a few turning points, the estimate may be improved by adding more information such as the line of sight velocities or the distances along the stream if they are available. In our tests with simulated streams, we did not find cases where the fitting procedure yielded significantly incorrect estimates of the input halo flattening (though of course with insufficient information there are degenerate solutions). Exhaustive testing is required to confirm that this is always the case. In all our tests, we used low mass satellites ($M  \leq 1.0 \times 10^{8} \msun$). This method cannot be applied in its current form to heavy and highly disrupted streams such as the Giant Stellar stream around M31 (\citealt{Ibata:2001p4924}). However, it may be possible to further extend the technique to such massive satellites by accounting for dynamical friction as an additional correction. 

We find that the line of sight distance of the progenitor from the center of the host galaxy is well-estimated (at times, with a sign discrepancy) when only the projection of the stream on the sky is available. The other orbital parameters, i.e. the velocity components of the progenitor are also well-estimated, but only when kinematic information, in the form of line of sight velocities along the stream or the \ion{H}{1} rotational curve, are provided. 

We also fit the streams generated in a spheroidal density model of the halo with orbits in a logarithmic potential. We see that despite using a simpler model, we are able to get a good estimate on the flattening of the halo. In the tests with the orbits in a logarithmic potential, we see that the initial distance $y_{0}$ is also constrained very well, but not the circular velocity. This is as expected; even for a circular orbit in a Keplerian potential, it is not possible to deduce the central mass from the radius of the orbit alone. The method cannot be used to constrain the inner slope of a spheroidal halo as the considered halo streams are too distant to probe the inner mass distribution using only the projection of the stream. It is encouraging to note that with circular velocities and the projection of the stream, most of the parameters of the halo can be recovered, as such information is available or easily acquirable for several systems. This technique is readily applicable to many of the streams observed in the near universe such as the low mass streams around M31, NGC 5907, NGC 891 and the Sagittarius stream in our galaxy.

\section*{Acknowledgments}
We are grateful to Peter Teuben and Walter Dehnen for their work on Nemo Stellar Dynamics Toolbox, GalPot and \emph{gyrfalc(ON)}, which have been the primary tools of our simulations and tests, as well as for their feedback on issues related to the same. We would also like to thank Jorge Pe\~{n}arrubia and Mustapha Mouhcine for helpful discussions during the development of the methodology. We are also very grateful to the reviewer, Dr. Shoko Jin, for insightful and detailed comments that helped improve the clarity of the paper. RI acknowledges support from the Agence Nationale de la Recherche (programme POMMME). GFL gratefully acknowledges the Australian Research Council for support through DP0665574 and DP110100678, and for his Future Fellowship (FT100100268). 

\begin{table*}
\caption{Models used in simulations: The parameters of the halo except its flattening, and the baryonic components of the host galaxy in which the streams are generated, along with their units are given. $\rho_{0}, r_{0}, q_{\rho}, \gamma, \beta$ and $r_{t}$ are the central density, scale radius, density flattening, inner power slope, outer power slope and truncation radius of the spheroidal components (the bulge and the halo). The parameters of the disk components are the central surface density $\Sigma_{d}$, the scalelength $R_{d}$, the scaleheight $z_{d}$ and the central depression parameter $R_{m}$ (which is 0 for the thin and thick disks).}
\begin{center}
\begin{tabular}{| l | l | l | l | l |}
 \hline
 Component & Parameter & Model 1& Model 2\\ 
                        &  & Streams A, D, G, C, F, I & Streams B, E, H\\ \hline
 \multirow{5}{*}{Halo} & $\rho_{0}$ ($\msun pc^{-3}$) & $0.11013117$ & $ 2.7986274$\\
 &  $r_{0}$ (kpc)  & 5.23553 & 1.0\\
 & $\gamma$ & 1.0 & 1.28 \\
 & $\beta$ & 3.0 & 2.34\\
 & $r_{t}$ (kpc) & 1000 & 1000\\ \hline
 \multirow{6}{*}{Bulge} & $\rho_{0}$ ($\msun pc^{-3}$) & $ 0.65074246$ & $0.3$ \\
 &  $r_{0}$  (kpc) & 1.0 & 1.0\\
 & $q_{\rho}$ & 0.6 & 0.6\\
 & $\gamma$ & 1.8 & 1.8\\
 & $\beta$ & 1.8 & 1.8\\
 & $r_{t}$ (kpc) & 1.9 & 1.9\\ \hline
 \multirow{3}{*}{Thin Disk} & $\Sigma_{d}$ ($\msun pc^{-2}$) & $412.0$  & $ 341.0$\\
 & $R_{d}$ (\kpc) & 3.2 & 3.2\\
 & $z_{d}$ (\kpc) & 0.18 & 0.18 \\  \hline
 \multirow{4}{*}{ISM} & $\Sigma_{d}$ ($\msun pc^{-2}$) & $69.5$  & $57.5$\\
 & $R_{d}$ (kpc) & 6.4 & 6.4\\
 & $z_{d}$ (kpc) & 0.04 & 0.04\\  
 & $R_{m}$ (kpc) & 4.0 & 4.0\\ \hline
\multirow{3}{*}{Thick Disk} & $\Sigma_{d}$ ($\msun pc^{-2}$) & $29.5$ & $24.4$\\
 & $R_{d}$ (kpc) & 3.2 & 3.2\\
 & $z_{d}$ (kpc) & 1.0 & 1.0\\  

\hline
\end{tabular}
\end{center}
\label{tablemodels}
\end{table*}%

\begin{table*}
\caption{Test Streams: Orbital Parameters, Halo Flattening and Age. $x_{0}, y_{0}, z_{0}$ are the initial positions and $v_{x_{0}}, v_{y_{0}}, v_ {z_{0}}$ the initial velocities of the orbits of the King model satellites of masses $m_{sat}$, tidal radii $r_{t}$ and central potential $\Phi(0)/\sigma^{2}$. The density flattening of the halo $q_{\rho}$ in each case is given in the first column. The ages given are the time since the starting of the simulation which gives the configurations of the streams that are used in the fitting.}
\begin{center}
\begin{tabular}{| l | l l l l | l l l | l l l l |}
\hline
Stream & $q_{\rho}$ & $x_{0}$ & $y_{0}$ & $z_{0}$ & $v_{x_{0}}$ & $v_{y_{0}} $ & $v_{z_{0}}$ & $Age$ & $m_{sat}$ & $r_{t}$ & $\Phi(0)/\sigma^{2}$\\ 
             &         & ($\kpc$) & ($\kpc$) & ($\kpc$) & ($\kms$) & ($\kms$) & ($\kms$) & (Gyr) & ($\times 10^{7} \msun$) & ($\kpc$) &\\ \hline
 A & 0.8 &  30.0 & 15.0 & 38.0 & -110.0 & 60.0 & -90.0 & 7.0 & 0.5 & 0.8 & 4.0\\
 B & 0.5 & 35.0 & 5.0 & 35.0 & -100.0 & 5.0 & 80.0 & 8.0 & 0.5 & 0.8 & 4.0\\
 C & 1.6 & 35.0 & 20.0 & 30.0 & -90.0 & 80.0 & 140.0 & 9.0 & 0.5 & 0.8 & 4.0\\
 \hline
 D & 1.0 & 35.0 & 10.0 & 35.0 & -120.0 & 30.0 & 100.0 & 10.0 & 5.0 & 1.2 & 4.0 \\
 E & 0.3 & 35.0 & 5.0 & 35.0 & -100.0 & 5.0 & 80.0 & 5.0 & 5.0 & 1.5 & 4.0\\
 F & 1.8 & 35.0 & 10.0 & 35.0 & -120.0 & 30.0 & 100.0 & 4.0 & 5.0 & 1.2 & 4.0\\
 \hline
 G & 1.0 & 25.0 & 20.0 & 20.0 & -120.0 & 90.0 & 140.0 & 8.0 & 10.0 & 1.7 & 7.0\\ 
 H & 0.5 & 35.0 & 5.0 & 35.0 & -100.0 & 5.0 & 80.0 & 8.0 & 10.0 & 1.1 & 2.75\\ 
 I & 2.0 & 30.0 & 0.0 & 0.0 & 180.0 & 160.0 & 130.0 & 7.0 & 10.0 & 0.9 & 3.0\\ 
\hline
\end{tabular}
\end{center}
\label{tablestreams}
\end{table*}

\bibliography{paper1_astroph}
\bibliographystyle{mn2e}

\end{document}